\documentclass[11pt]{article}

\usepackage[utf8]{inputenc}
\usepackage{amsmath,amssymb,amsfonts}
\usepackage{graphicx}
\usepackage{geometry}
\usepackage{cite}
\usepackage{authblk}
\usepackage{hyperref}
\usepackage{multicol}
\usepackage{caption}
\usepackage{indentfirst}
\usepackage{bm}
\usepackage{float}
\usepackage{tablefootnote}
\usepackage{threeparttable}
\usepackage{url}
\usepackage{tabularx}
\usepackage{makecell}
\usepackage{array}
\usepackage{multirow}

\title{\textbf{Neural Refractive Index Primitives for Flame Field Reconstruction Using Background-Oriented Schlieren}}

\author[1,2]{Xinyi Lu}
\author[1,2,*]{Wei Hu}
\author[2]{Zizhou Liao}
\author[2]{Zheng Wang}
\author[2]{Yue Zhang}
\author[1,2,*]{Jingxuan Li}

\affil[1]{Aircraft and Propulsion Laboratory, Ningbo Institute of Technology, Beihang University, Ningbo 315800, China}
\affil[2]{School of Astronautics, Beihang University, Beijing 100191, China}
\affil[*]{\textit{Corresponding authors:} \texttt{weihu22@buaa.edu.cn}; \texttt{jingxuanli@buaa.edu.cn}}

\date{}

\begin{document}

\maketitle
\vspace{-5em}
\begin{center}
\url{https://github.com/Weihu22/Neural-Implicit-Reconstruction-for-BOS}
\end{center}
% ===== 单栏：Abstract =====
\noindent\textbf{Abstract}

An improved neural refractive-index-primitive method for background-oriented schlieren tomography is presented, enabling continuous three-dimensional reconstruction of refractive-index fields using a compact multilayer perceptron. The method adopts the refractive-index field as the sole neural primitive and integrates multiresolution hash encoding, automatic-discrete gradient losses, and a three-dimensional mask to enable fast convergence and high-resolution, spatially coherent reconstructions. Tests on numerical combustion phantoms and real flame data demonstrate accurate recovery of both large-scale structures and fine-scale turbulence, strong robustness to noise, and clear advantages over frequency-encoding-based and voxel-based reconstruction methods.

\vspace{0.5em}

\vspace{1em}

% ===== 开始双栏 =====
\begin{multicols}{2}

\noindent\textbf{Novelty and significance statement}
\indent This work presents a novel neural framework for reconstructing three-dimensional refractive-index fields from background-oriented schlieren measurements of flames. By integrating different encoding strategies, various gradient computations, a 3D flow mask, and a ray-sampling scheme, the method enables fast high-fidelity reconstruction with a compact network. The proposed framework improves optimization stability and convergence speed, enabling more accurate refractive-index reconstruction and providing a more reliable basis for temperature inference under restricted assumptions in air-diluted or well-mixed combustion regimes.

\section{Introduction}

Combustion measurements involve spatial scales spanning several orders of magnitude in both laboratory and propulsion systems. As flames grow from millimeter to meter scales, full-field diagnostics become increasingly limited. Contact sensors and many optical methods are constrained by measurement range and complex setups, making it difficult to resolve unsteady, irregular three-dimensional (3D) flow fields over large domains.

Compared with these traditional diagnostic techniques, background-oriented schlieren (BOS) provides a large field of view and a relatively simple hardware setup, making it well suited for large-area measurements in combustion processes \cite{raffel2015}
\cite{schmidt2025}. BOS relies on the deflection of light rays as they traverse an inhomogeneous refractive-index field. The ray-deflection angle is obtained through geometric optics by first estimating the displacement between the reference and disturbed background images via image processing. The refractive-index field is then reconstructed via inverse ray equation and subsequently converted into density and temperature fields using the Gladstone–Dale relation and the ideal-gas equation of state \cite{grauer2018}. Notably, BOS measures refractive index variations rather than temperature directly, and temperature inference typically requires additional assumptions (e.g., constant mixture properties) or more advanced data assimilation techniques \cite{martins2023}\cite{zheng2025}.

For general, irregular, and inhomogeneous flow fields, BOS must be combined with tomographic methods to reconstruct the refractive-index distribution from displacement data acquired at multiple viewing angles. A typical BOS tomography system deploys several cameras (often ten or more) around the flame to acquire multi-view projections \cite{cai2022}. Compared with analytical reconstruction methods such as filtered back-projection, a general class of algebraic reconstruction techniques are widely used in BOS tomography due to their flexibility in handling sparse and irregular measurements \cite{liu2021}. In the classical algebraic reconstruction technique (ART), originally proposed by Gordon and Herman, the ray equation is discretized into a linear system of the form Ax = b, and the solution is updated sequentially using a Kaczmarz-type projection algorithm \cite{gordon1970}. In practice, its stability is often achieved through early termination of iterations, which acts as an implicit form of regularization.

Beyond classical ART, other algebraic approaches have been developed based on different discretization and inversion strategies. For example, Atcheson et al. discretized various hot-air and flame flows into voxels and employed radially symmetric basis functions combined with visual hull masking to recover three-dimensional refractive-index gradients, which were then integrated to obtain the refractive-index and volume density fields \cite{atcheson2008}. Unlike such two-step approach, Nicolas et al. proposed a direct reconstruction method based on a GPU-accelerated conjugate-gradient solver, demonstrating instantaneous density reconstructions on grids up to 130 megavoxels from both simulations and experimental datasets \cite{nicolas2016}.

Because of the large number of unknowns in the flow field, additional constraints are required to ensure a unique and physically meaningful solution. Voxel-based algebraic solvers typically impose global smoothness priors, such as total variation or Tikhonov regularization \cite{nicolas2016}. However, these simple constraints cannot preserve fine-scale physical structures, including multi-scale turbulence. This limitation is exacerbated in dense voxel-based reconstructions of high-resolution combustion fields, where global smoothness constraints cause excessive smoothing, blurring, and structural distortions.

In recent years, neural implicit reconstruction techniques (NIRT) have advanced BOS tomography due to their strong feature extraction and nonlinear mapping capabilities. In NIRT, coordinate-based neural networks map spatial (or spatiotemporal) coordinates to physical fields such as refractive index, density, or temperature for BOS reconstruction \cite{molnar2025}.

One of the earliest and most representative NIRT approaches for BOS uses physics-informed neural networks (PINNs) to incorporate governing physical laws with experimental measurements, enabling inference of continuous three-dimensional temperature, velocity, and pressure fields from space–time data alone \cite{cai2021pinn}. However, early formulations relied on mismatch losses against conventional algebraic reconstructions, limiting their accuracy by the quality of the underlying solvers.

To overcome this limitation, a major improvement in NIRT involved replacing the field-to-field mismatch loss with a loss defined between predicted and measured projections, thereby transforming the training process from supervised to unsupervised learning and significantly enhancing model generalization. For example, Monlar et al. used a deflection-angle mismatch loss in their PINN mode to simultaneously reconstruct density, velocity, and pressure fields in axisymmetric underexpanded jets \cite{molnar2023}. However, since NIRT takes low-frequency spatial and temporal coordinates as its inputs, the unsupervised training process often becomes unstable or underfitting when dealing with complex flow structures or noisy measurements.

Another key advancement in NIRT is re-encoding low-frequency coordinates to enhance the network’s ability to represent high-frequency structures. The most commonly used method is frequency encoding, also known as Fourier encoding, which maps smooth, low-frequency physical space and time coordinates into a high-dimensional feature space constructed from multi-scale sinusoidal and cosine functions. This explicit injection of high-frequency information improves the network’s capacity to represent sharp gradients and fine-scale structures in flame flow fields. The frequency encoding strategy has been employed in many works to reconstruct density fields in laminar flames with irregular structures and in supersonic gas jets \cite{zheng2025}\cite{molnar2025}\cite{he2025}.

In fact, NIRT for BOS shares strong conceptual similarities with the Neural Radiance Field (NeRF) framework widely used in computer vision \cite{mildenhall2022}\cite{muller2022}. Both represent target fields through coordinate-based neural networks. Their primary difference lies in the physical rendering model: NIRT integrates refractive-index gradients along rays, while NeRF integrates radiance contributions resulting from reflection, scattering, and absorption processes along the ray path. Compared to NIRT, NeRF has undergone rapid development with a rich ecosystem of variants and highly efficient encoding strategies, providing a mature methodological foundation for advancing BOS inversion techniques. For example, Li et al. adopted the hash encoding strategy to reconstruct the refractive-index gradient field of swirling flames with complex spatial distributions using a compact network, without incorporating physical equation constraints \cite{li2024}. Notably, their method directly reconstructs refractive-index gradients rather than the refractive index itself, thereby omitting the step of deriving the index field from its gradients. Compared with directly reconstructing the gradient field, simultaneously predicting both the gradients and the refractive-index field increases network convergence difficulty and exacerbates the ill-posedness of the reconstruction.

The aim of this work is to develop a more compact and faster-converging neural implicit reconstruction network to improve the accuracy of refractive-index field reconstruction, which may in turn support temperature reconstruction only under restricted assumptions. In particular, in certain highly air-diluted or well-mixed combustion configurations, where $N_2$ remains dominant and composition-induced variations are comparatively modest, refractive-index variations may be approximated as being primarily temperature-driven. 

In this study, the refractive-index field is used as the sole neural primitive within a unified reconstruction framework that incorporates an encoding strategy, a field gradient loss, three-dimensional mask constraints, and a ray-sampling scheme. By jointly enforcing these complementary components, the proposed formulation maintains a simple model structure while promoting stable convergence and accurate reconstruction. We further evaluate the effects of encoding, gradient formulation, and mask design on temperature reconstruction using both synthetic and experimental flames across varying turbulence intensities.

\section{Methodology}

\subsection{Framework}
\begin{figure}
    \centering
    \includegraphics[width=0.8\linewidth]{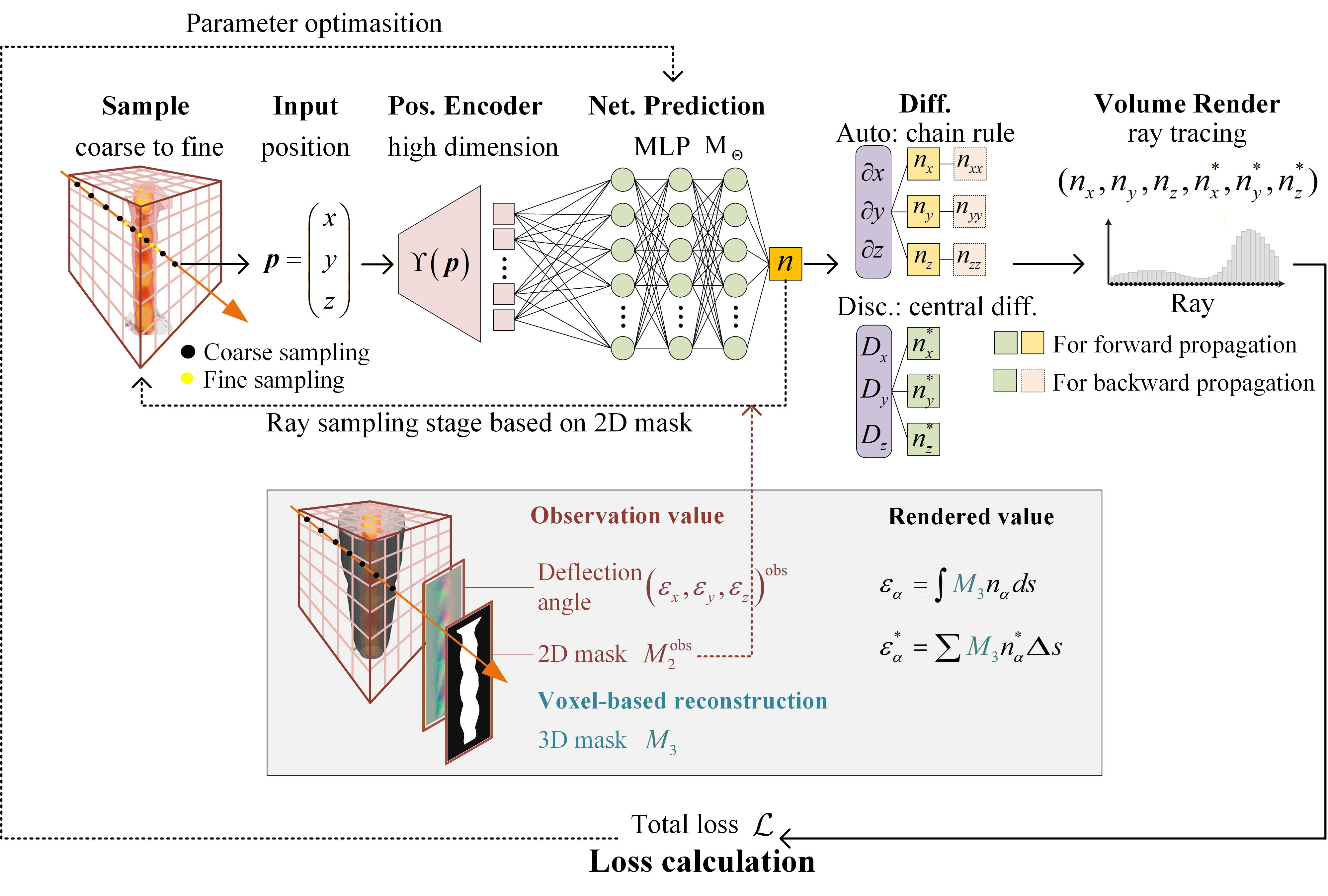}
    \caption{An overview of our neural refractive-index-primitive framework for background-oriented schlieren tomography.}
    \label{fig:frame}
\end{figure}

Fig.\ref{fig:frame} visualizes the overall pipeline of our neural refractive-index-primitive framework for background-oriented schlieren tomography. We represent an instantaneous flow field as a continuous 3D function that outputs the refractive index (n) at each spatial location ${\bf{p}} = (x,y,z)$. This function is parameterized by a fully connected deep neural network without any convolutional layers—commonly referred to as a multilayer perceptron (MLP). The network is trained to regress from a single 3D coordinate ${\bf{p}} = (x,y,z)$ to the corresponding refractive index value.

To render this neural refractive-index field (NeRIF) from a particular viewpoint, we proceed as follows:\\
1) The camera rays are marched through the refractive-index field toward the background plane to obtain a set of sampled 3D points along each ray;\\
2) The sampled 3D coordinates are encoded using a positional encoding that enables the MLP to represent high-frequency spatial variations;\\
3) The encoded coordinates are fed into the network to predict the refractive index values at those locations;\\
4) The classical ray-tracing principles are applied to accumulate the spatial derivatives of the refractive index $({n_x},{n_y},{n_z})$ along each ray, yielding a 3D deflection vector $({\varepsilon _x},{\varepsilon _y},{\varepsilon _z})$.

\end{multicols}

\vspace{-2em}
\begin{multicols}{2}
Since the refractive index is modeled as a continuous differentiable function, we use automatic differentiation to compute gradients  accurately and efficiently. These gradients are themselves differentiable, enabling optimization of all network parameters via gradient descent. The model is trained by minimizing the discrepancy between observed deflection angles, obtained from optical flow between background-image pairs, and the corresponding deflections rendered from the neural representation. Minimizing this error across multiple viewpoints enforces multi-view consistency and encourages the network to predict a physically plausible, spatially coherent refractive-index field that best explains the BOS observations.

We observe that a basic NeRIF optimization struggles to converge to complex combustion fields. To address this, we investigate several key design components within a unified framework:\\
1) Encoding strategy. We compare multiresolution hash encoding with traditional frequency encoding in terms of their ability to represent complex flow structures and reconstruction performance.\\
2) Gradient formulation. Automatic, discrete, and combined automatic–discrete formulations for refractive-index gradient computation are considered, and their effects on convergence behavior and reconstruction stability are evaluated.\\
3) Mask and sampling design. A 2D mask extracted from BOS observations is used to guide the ray-sampling scheme, while a corresponding 3D mask imposes hard constraints on the reconstruction domain. The impact of the 3D mask is further evaluated through ablation studies. 

\subsection{Neural refractive index primitives}
\subsubsection{Position encoding}

Frequency encoding maps positions $\bm{p}$ in a low dimensional space $\mathbb{R}$ into a higher-dimensional feature space $\mathbb{R}^{2N_\mathrm{f}}$ by introducing multi-scale sine and cosine functions, thereby enhancing the model’s ability to represent high-frequency details. The input coordinates are typically transformed using a set of periodic functions with different frequencies, as defined below:

\begin{equation}
\begin{aligned}
\left(
\sin\left(2^0 \pi \bm{p} \right),\,
\cos\left(2^0 \pi \bm{p} \right), \ldots, \right. \\
\left.
\sin\left(2^{N_\mathrm{f} - 1} \pi \bm{p} \right),\,
\cos\left(2^{N_\mathrm{f} - 1} \pi \bm{p} \right)
\right)
\end{aligned}
\end{equation}

\begin{figure}[H]
    \centering
    \includegraphics[width=1\linewidth]{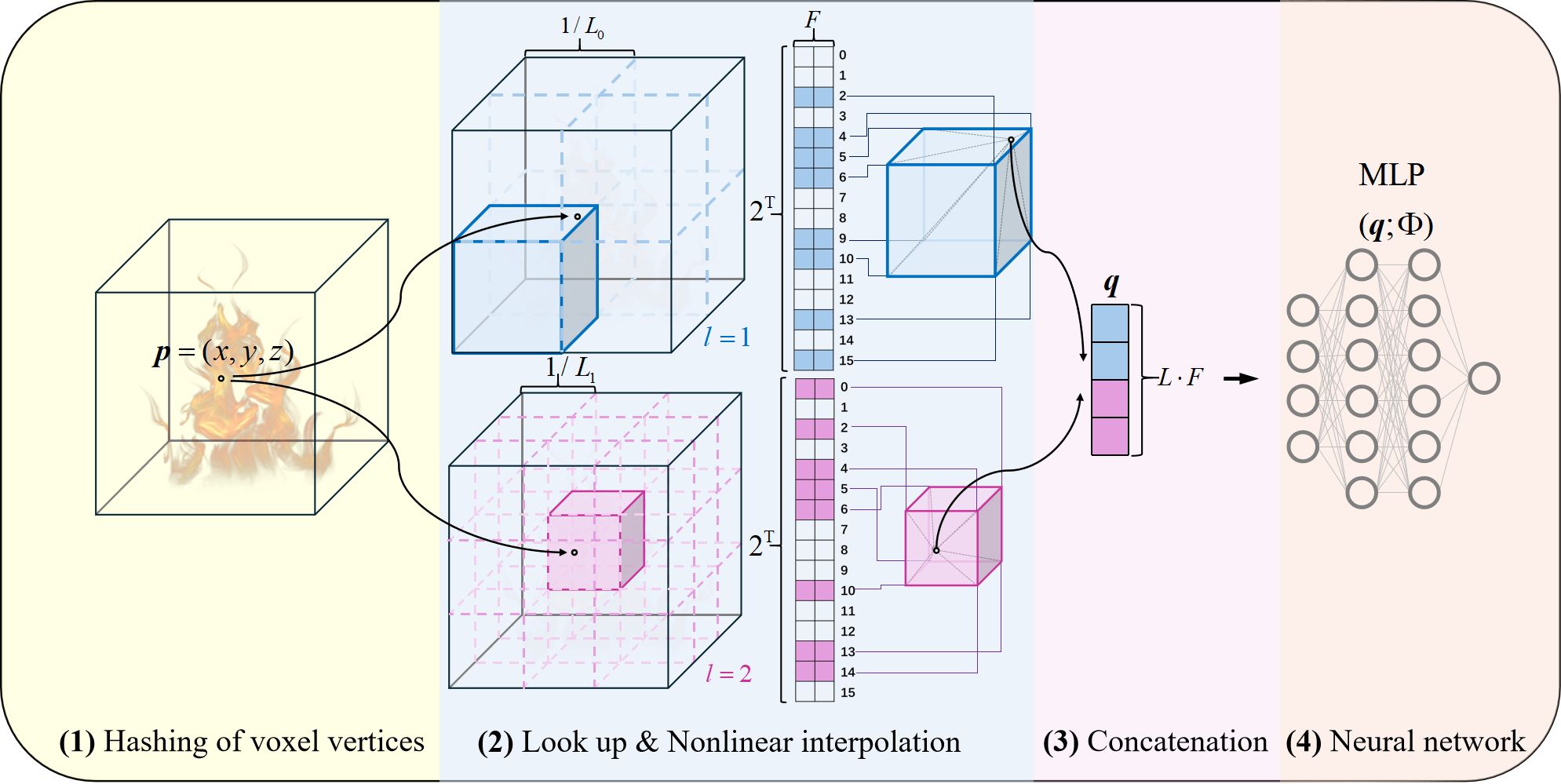}
    \caption{The overview of the multi-resolution hash encoding scheme functioning in 3D coordinate system.}
    \label{fig:hash}
\end{figure}

Fig.\ref{fig:hash} illustrates the multi-resolution hash encoding scheme. It maps continuous 3D coordinates into a hierarchy of $L$ levels of learnable hash tables at different resolutions. Each level has a learnable embedding table containing $2^{T}$ entries, each storing an $F$-dimensional feature vector. For a given input coordinate, we identify the eight enclosing voxel vertices at each level and map their indices to the hash table using a hash function.

Subsequently, the retrieved embedding features of the eight vertices are fused using cubic nonlinear interpolation, where the interpolation weights are used to compute a weighted average of the vertex features. Finally, the interpolated features from all levels are concatenated along the feature dimension to form the resulting multi-scale spatial representation:

\begin{equation}
[{f_1}({\bm{p}}),{f_2}({\bm{p}}), \ldots ,{f_L}({\bm{p}})],
\end{equation}
where $f_i(\bm{p})$ represents the feature vector obtained by applying nonlinear interpolation to the vertex features of the $i$-th grid level that contains the point $\bm{p}$. $L$ denotes the number of resolution levels in the multi-resolution hash encoding. For each query point, interpolated feature vectors from all levels are concatenated into a single output vector, forming a multi-scale representation that captures both global structure and fine details. This feature vector is then fed into the subsequent neural network module for refractive-index prediction.

\end{multicols}
\vspace{-1.5em}
\begin{multicols}{2}

During backpropagation, the hash-table feature vectors are treated as learnable parameters and updated via gradient descent to minimize reconstruction error. Although the hash indexing involves discrete integer operations and is not differentiable, the embedding vectors retrieved from the hash table are optimized directly. Gradients are propagated through the differentiable interpolation step and accumulated on the corresponding embedding entries. 

In BOS measurements, the observations correspond to the line-integrated spatial gradient of the refractive index rather than the field itself. While BOS deflections depend only on the first spatial derivatives of the refractive index, gradient-based optimization with automatic differentiation requires differentiating these gradients during backpropagation, which introduces second-order spatial derivatives in the computational graph However, standard hash encoding employs linear interpolation, whose piecewise-linear form yields zero second-order derivatives and is therefore incompatible. To overcome this limitation, we replace linear interpolation with a cubic nonlinear interpolant. Specifically, we adopt a smoothstep interpolant, which provides non-zero second-order derivatives and ensures that first derivatives smoothly decay to zero at interval boundaries, improving continuity and optimization stability.

\subsubsection{Volume rendering for refractive index induced deflection}
Our 3D NeRIF models the scene by assigning a volumetric refractive index to every point in space. We compute the refractive-index gradient encountered by any ray traversing the scene using linear ray tracing under the paraxial approximation. For a camera ray parameterized as ${\bf{r}}\left( t \right) = {\rm{ }}{\bf{o}} + t{\bf{d}}$ with near and far bounds ${t_{\rm{n}}}$ and ${t_{\rm{f}}}$, the expected deflection angle is given by:
\begin{equation}
\varepsilon_{\alpha} = \frac{1}{n_0} \int_{t_n}^{t_f} \frac{\partial n}{\partial \alpha} \, dt, 
\quad \alpha \in \{x, y, z\}
\end{equation}
where ${\varepsilon _\alpha }$ is the deflection angle along the $\alpha$-axis direction,  is the ambient refractive index, and $n_0$ represents the local refractive index.

We numerically estimate the continuous integral using discrete quadrature. To improve sampling efficiency, we adopt a multi-resolution hierarchical sampling strategy \cite{muller2022}, where the sampling step along each ray is dynamically adjusted based on the predicted refractive index. The core idea of this approach is twofold: empty space can be skipped, while regions with high refractive index are sampled more densely using small, uniform steps.

\begin{figure}[H]
    \centering
    \includegraphics[width=1\linewidth]{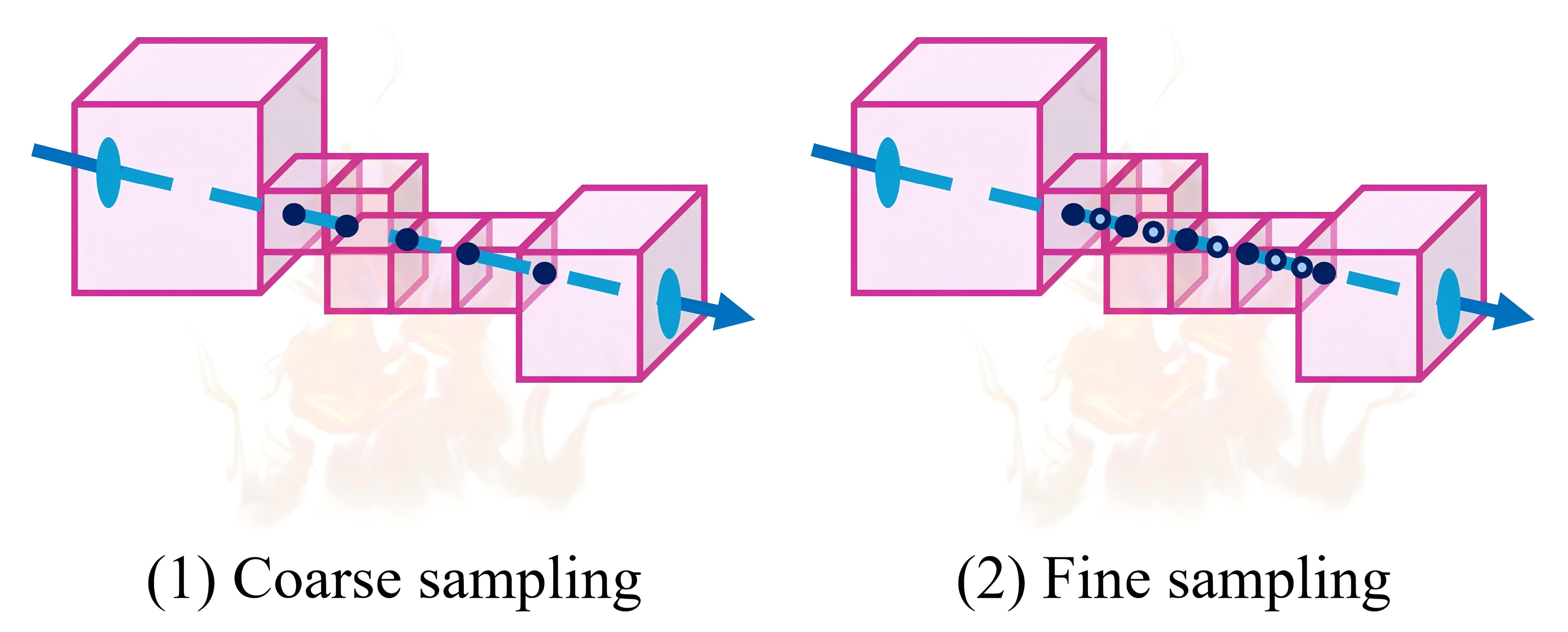}
    \caption{Sampling strategy of multiscale occupancy grids.}
    \label{fig:sampling}
\end{figure}

This is achieved using multiscale occupancy grids that distinguish empty from non-empty regions, as shown in Fig.\ref{fig:sampling}. Each voxel stores a single-bit occupancy value, and samples below a threshold are skipped. These grids are independent of the neural encoding and are updated during training based on the absolute refractive-index magnitude. The refractive index is pre-normalized so that the ambient value is zero (Section \ref{sec:Implementation}), and regions with larger deviations are sampled more densely due to stronger ray deflection.

Furthermore, to reduce aliasing from fixed grid-aligned sampling, small stochastic perturbations are introduced into the ray sampling process. A random offset is applied to each ray’s initial sampling position to avoid identical starting depths. In addition, voxel-level jitter is used when querying occupancy so that sub-voxel locations are sampled across iterations rather than repeatedly accessing voxel centers. These perturbations decorrelate sampling from the grid structure, improving the robustness and accuracy of hierarchical sampling.

Consequently, the rendering function of Equation \eqref{eq:epsilon_discrete} is discretized as follows:
\begin{equation}
\varepsilon_{\alpha} \approx \sum_{i=1}^{K} 
\frac{\partial \sigma(\bm{p}_i)}{\partial \alpha} \, \Delta t_i
\label{eq:epsilon_discrete}
\end{equation}
where $K$ is the maximum number of sampling steps along a single ray, $\Delta {t_i}$ is the step size at the $\rm{i}$-th sample point, and $\sigma ({{\bm{p}}_i})$ denotes the refractive-index field evaluated at the spatial location ${{\bm{p}}_i}$.

\subsubsection{Loss}
\label{sec:Loss}
The model is optimized by minimizing the discrepancy between the observed deflection angles and those produced by our renderer. To compute the rendered deflection, we systematically evaluate automatic, discrete spatial, and combined (automatic–discrete) gradient formulations of the predicted refractive index. The resulting deflection is compared with observations, and the loss is defined as a mean squared error (MSE) between them:
\begin{equation}
\mathcal{L} = \frac{1}{N} \sum_{i=1}^{N} 
\left(
\lambda_a \left\| \varepsilon_i - \varepsilon_i^{\mathrm{obs}} \right\|_2^2
+ \lambda_d \left\| \varepsilon_i^{*} - \varepsilon_i^{\mathrm{obs}} \right\|_2^2
\right)
\label{eq:loss}
\end{equation}
where $N$ is the number of sampled rays, $\varepsilon$ (${\varepsilon ^ * }$) denotes the deflection angles computed from automatic (discrete) spatial gradients, and ${\varepsilon ^{{\rm{obs}}}}$ is the observed deflection angle obtained from optical flow between pairs of background images. ${\lambda _a},{\lambda _d} \in \left\{ {0,1} \right\}$ indicates gradient type.

For each position ${\bm{p}} = (x,y,z)$, the spatial gradient of the normalized refractive index is computed using the automatic differentiation formula:
\vspace{-0.5em}
\begin{equation}
\nabla \sigma(\bm{p}) =
\left(
\frac{\partial \sigma}{\partial x},\;
\frac{\partial \sigma}{\partial y},\;
\frac{\partial \sigma}{\partial z}
\right)
\label{eq:grad_sigma}
\end{equation}
The corresponding discrete gradient is computed using the central difference method:
\begin{equation}
\nabla \sigma(\bm{p})^{*} \approx 
\frac{\sigma(\bm{p} + \Delta \bm{l}) - \sigma(\bm{p} - \Delta \bm{l})}{2\,\Delta}
\label{eq:grad_fd}
\end{equation}
where $l$ represents the unit vectors in the $x$, $y$, and $z$ directions, $\Delta  = {b \mathord{\left/
 {\vphantom {b {\left( {2K} \right)}}} \right.
 \kern-\nulldelimiterspace} {\left( {2K} \right)}}$ represents their magnitude, $b$ is the minimum side length of the region of interest in the flow field. 

\subsection{Combination with field mask}
Since real flow fields occupy a finite region, masking is used to constrain the reconstruction domain, avoiding inversion in irrelevant areas (e.g., solids or ambient air) and reducing unnecessary sampling. Molnar \& Grauer et al. incorporated a 2D mask into the deflection-angle loss for temporal BOS tomography [11], suppressing background or occluded pixels to focus optimization on valid flow regions, thereby improving stability and efficiency.
\vspace{-1em}
\begin{figure}[H]
    \centering
    \includegraphics[width=0.9\linewidth]{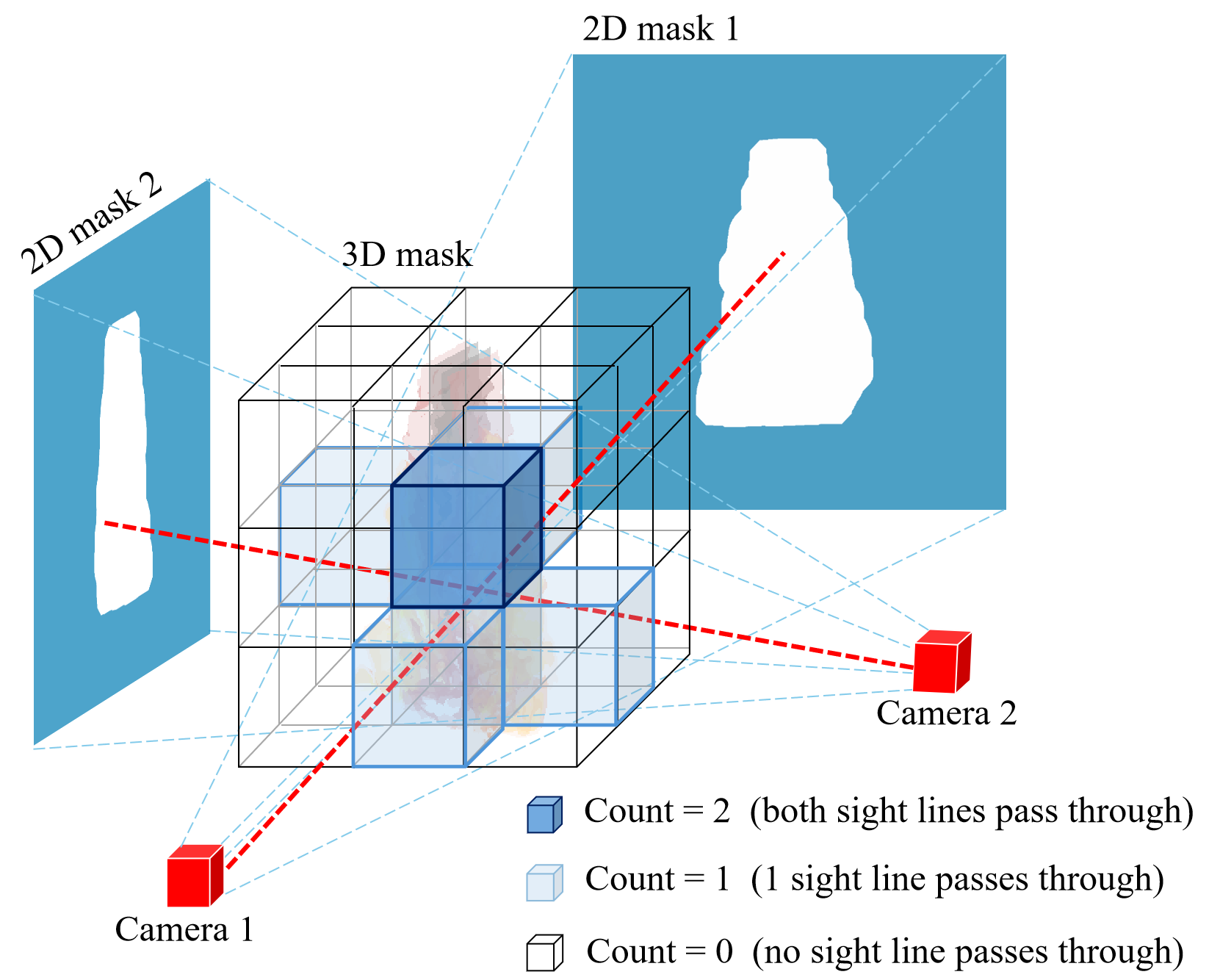}
    \caption{2D masks and 3D mask.}
    \label{fig:mask}
\end{figure}
\vspace{-1.5em}
However, a 2D mask only constrains projection-space data and cannot localize occlusions in the 3D volume, allowing spurious refractive-index values to persist. In contrast, a 3D mask directly constrains the volume, enabling rays to skip invalid voxels, reducing computation and suppressing noise. Nicolas et al. applied a volumetric support constraint in direct BOS 3D density inversion, setting zero values outside the support to suppress artifacts and improve stability \cite{nicolas2016}. Li et al., introduced a posterior support constraint that removes voxels with nearly constant refractive index, reducing ineffective swirling flames regions and enhancing volumetric reconstruction quality \cite{li2024}.
Here, we incorporates a 3D mask into neural refractive-index primitives for BOS tomography. Together with the 2D mask used for ray sampling, this forms a complementary two-level masking strategy, where the 3D mask constrains the reconstruction domain and the 2D mask guides ray sampling. The implementation pipeline is shown in Fig.\ref{fig:mask}.
\vspace{-1em}
\subsubsection{2D mask preprocessing}
\vspace{-0.5em}
First, a 2D mask is generated from the observed deflection-angle data. The squared deflection magnitudes are normalized and thresholded to form an initial binary mask. The threshold is set as the minimum value that preserves the displacement signal while avoiding background over-extension. To improve boundary continuity and fill local holes, morphological dilation with a circular structuring element is applied, yielding the final 2D binary mask ${M_2}\left( {u,v} \right)$, where 1 denotes valid regions and 0 denotes invalid regions. Only pixels in the valid regions are used to generate training rays.
\vspace{-1em}
\subsubsection{3D mask preprocessing}
\vspace{-0.5em}
Second, the 2D masks are back-projected into the voxel grid to define the valid 3D region. A voxel is marked valid (1) if it is observed by more than Nc cameras; otherwise, it is marked invalid (0).

In this work, Nc is typically set to the total number of cameras minus one, with smaller values yielding a more permissive mask and larger values enforcing stricter constraints. The resulting voxel-based 3D mask is interpolated into a continuous binary field$ {M_3}\left( {x,y,z} \right)$, where 1 denotes valid regions and 0 invalid ones. The predicted refractive index is then modulated by this mask to obtain a mask-gated field.
\section{Measurement scenarios}
\subsection{Dataset}
\vspace{-1.5em}
\begin{figure}[H]
    \centering
    \includegraphics[width=1\linewidth]{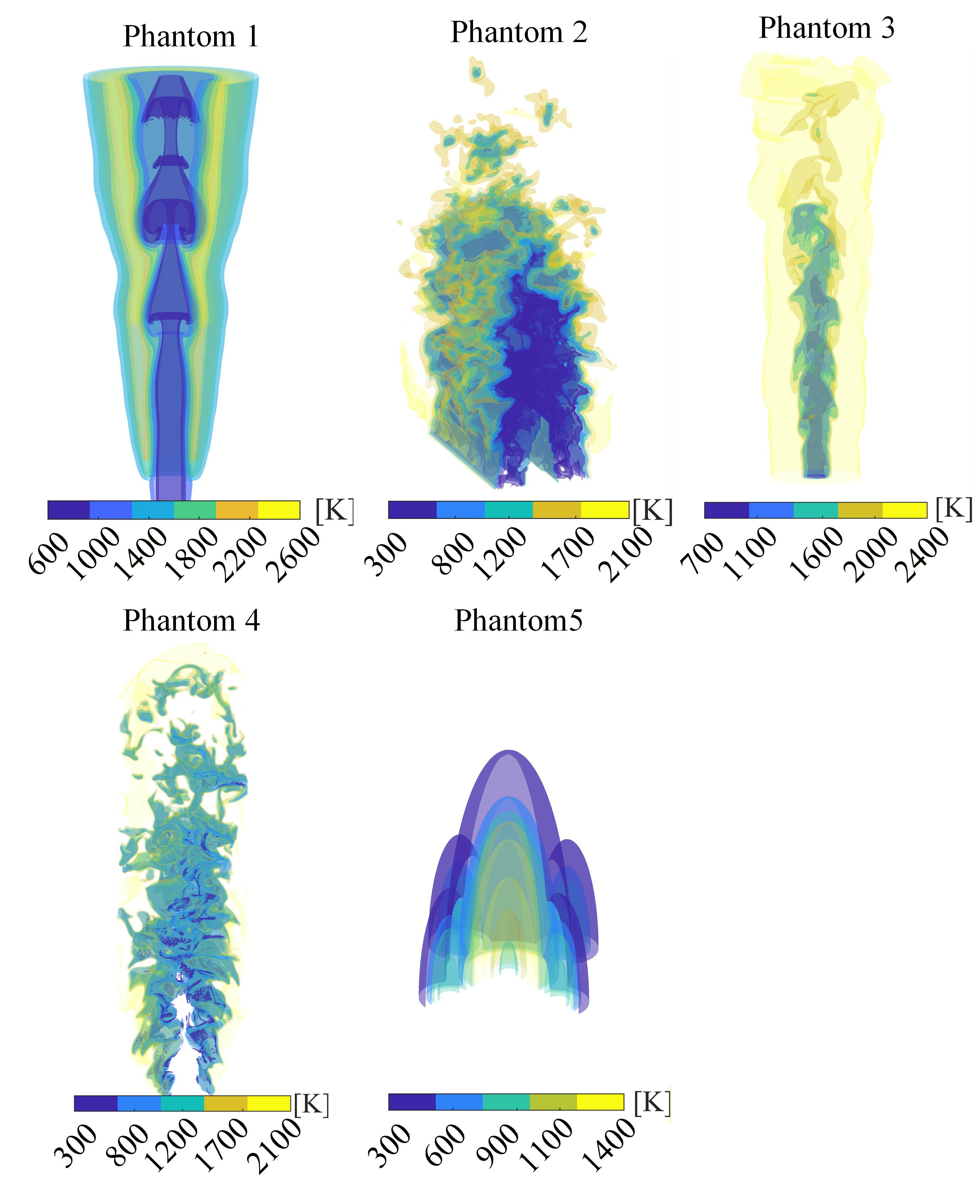}
    \caption{The isosurfaces of the true phantoms’ temperature.}
    \label{fig:dataset}
\end{figure}
\vspace{-1em}
In the numerical validation, five flame phantoms with different temperature distributions were used, including four high-fidelity cases from the BLASTnet dataset \cite{brouzet2021}-\cite{sharma2025} and one mathematical phantom \cite{cai2022}. Since the original BLASTnet data contain billions of voxels, a single snapshot from each case was selected and downsampled for efficient reconstruction. For Phantom 1, a 2D temperature field was extended to 3D under an axisymmetry assumption. Details are listed in Table \ref{tab:dataset}, with distributions shown in Fig.\ref{fig:dataset}.

In general combustion environments, composition variations make the refractive index–temperature mapping non-unique; however, since this work focuses on improving refractive-index reconstruction, and under restricted conditions (e.g., well-mixed, air-diluted flames where $N_2$ dominates) refractive-index variations can be approximated as primarily temperature-driven, we adopt a simplified relation to convert phantom temperature to refractive index\cite{qin2002}:
\begin{equation}
n = \frac{T_0 (n_0 - 1)}{T} + 1
\label{eq:refractive_index}
\end{equation}
where $T_0$ represents the ambient temperature (unit: K), $n_0$ represents the ambient refractive index, and $T$ corresponds to the phantom temperature.
\vspace{-1.5em}
\end{multicols}
\begin{table}[H]
\centering
\caption{Details of the phantoms and experimental flames.}
\label{tab:dataset}
\vspace{-0.5em}
\setlength{\tabcolsep}{3pt}

% ===== 表格主体（缩小）=====
{\fontsize{7pt}{8pt}\selectfont
\resizebox{\textwidth}{!}{
\begin{tabular}{c c c >{\raggedright\arraybackslash}p{3cm} p{5cm}}
\hline
No. & Fuel & Burner & \hspace{6mm}Ambient settings & \makecell{Resolution \\ (voxel size [$\mathrm{mm}^3$])} \\
\hline

1\textsuperscript{1} & H$_2$, N$_2$ & circular
& \makecell[l]{\hspace{8mm}$T_0=1100$ \\ \hspace{8mm}$n_0=1.00007253$}
& \makecell{$140\times294\times140$ \\ $(0.5\times0.5\times0.5)$ \\ $280\times588\times280$ \\ $(0.25\times0.25\times0.25)$} \\

2\textsuperscript{2} & H$_2$, air & slot
& \makecell[l]{\hspace{8mm}$T_0=2019$ \\ \hspace{8mm}$n_0=1.00003952$}
& \makecell{$80\times180\times80$ \\ $(0.9\times0.9\times0.9)$} \\

3\textsuperscript{3} & CH$_4$, air & round
& \makecell[l]{\hspace{8mm}$T_0=2413$ \\ \hspace{8mm}$n_0=1.00003307$}
& \makecell{$140\times240\times140$ \\ $(0.5\times0.5\times0.5)$} \\

4\textsuperscript{4} & NH$_3$, H$_2$, air & slot
& \makecell[l]{\hspace{8mm}$T_0=2160$ \\ \hspace{8mm}$n_0=1.00003694$}
& \makecell{$128\times512\times64$ \\ $(0.25\times0.25\times0.25)$} \\

5\textsuperscript{5}
& \multicolumn{2}{c}{\makecell{Formula: $T_5(x,y,z)$}}
& \makecell[l]{\hspace{8mm}$T_0=296.15$ \\ \hspace{8mm}$n_0=1.00027000$}
& \makecell{$144\times200\times144$ \\ $(0.5\times0.5\times0.5)$} \\

6 & C$_3$H$_8$, air & dual circular
& \makecell[l]{\hspace{8mm}$T_0=296.15$ \\ \hspace{8mm}$n_0=1.00027000$}
& \makecell{$176\times56\times176$ \\ $(0.25\times0.25\times0.25)$} \\

7 & C$_3$H$_8$, air & single circular
& \makecell[l]{\hspace{8mm}$T_0=296.15$ \\ \hspace{8mm}$n_0=1.00027000$}
& \makecell{$124\times188\times124$ \\ $(0.25\times0.25\times0.25)$} \\

\hline
\end{tabular}
}
}

\vspace{6pt}

% ===== Notes（单独控制）=====
\footnotesize
\begin{minipage}{\textwidth}
\textsuperscript{1} \url{https://blastnet.github.io/diluted_partially_premixed_h2air_lifted_flame} \\
\textsuperscript{2} \url{https://blastnet.github.io/premixed_slot_flame_h2air} \\
\textsuperscript{3} \url{https://blastnet.github.io/turbulent_round_jet_premixed_ch4air} \\
\textsuperscript{4} \url{https://blastnet.github.io/nh3h2air_premixed_slot_flame} \\
\textsuperscript{5}
{\scriptsize
$\textstyle
\begin{array}{l}
T_5(x,y,z) = 296.15 + 1196.85 e^{-\left( \tfrac{3z^2 + 3x^2}{20^2} + \tfrac{3(y - 50)^2}{80^2} - 0.9 \right)^2} 
+ 696.85 e^{-\left( \tfrac{3(z - 25)^2 + 3x^2}{20^2} + \tfrac{3(y - 50)^2}{80^2} - 0.9 \right)^2}\\
+ 696.85 e^{-\left( \tfrac{3(z + 25)^2 + 3x^2}{20^2} + \tfrac{3(y - 50)^2}{80^2} - 0.9 \right)^2} 
+ 696.85 e^{-\left( \tfrac{3(z - 25)^2 + 3x^2}{20^2} + \tfrac{3(y + 50)^2}{80^2} - 0.9 \right)^2}
+ 696.85 e^{-\left( \tfrac{3(z + 25)^2 + 3x^2}{20^2} + \tfrac{3(y + 50)^2}{80^2} - 0.9 \right)^2}
\end{array}
$
}
\end{minipage}

\end{table}

\begin{multicols}{2}

Twelve simulated cameras were evenly distributed over 165° at a radius of 800 mm from the sampling volume center. Each used a 105 mm lens and a 720×1280 sensor with 20 $\mathrm{\mu}$m pixel pitch. Backgrounds were placed 800 mm on the opposite side. In all synthetic cases, except for the robustness study in Fig. \ref{fig:noise}, the aperture number is fixed at 32, and noise-free conditions are assumed.  After defining the BOS system configuration and the refractive index distribution within the probe volume, a realistic deflection-sensing pipeline is implemented, explicitly simulating blur-circle formation, finite-aperture averaging via multi-ray integration, background warping, and optical-flow-based displacement estimation. This enables more faithful BOS image formation, including defocus blur, finite-aperture effects, and the indirect nature of displacement recovery.

Experimental validation was further conducted on two real flame cases, including a double flame and a diffusion flame. For the double flame case, the burner is equipped with a nozzle featuring two exits, each with a diameter of 8 mm and a height of 15 mm, producing two small flames. The center-to-center spacing between the flames is 10 mm. The fuel consists of air and C3H8 with a flow velocity of 1.38 m/s, corresponding to a Reynolds number of 796. For the diffusion flame case, the fuel flow velocity is 0.14 m/s with an exit diameter of 4 mm, resulting in a Reynolds number of 119, and the flame height is approximately 48.4 mm. 

The BOS experimental setup differs slightly between the two flame cases. For the double flame case, the steady flames allow multi-view acquisition by rotating the burner with an 18° step, yielding ten projection angles (0°–162°). A high-speed camera (NOVA s12, 1024×1024) with a 105 mm Nikon lens is used. Illumination is provided by a 60 W LED panel, with an exposure time of 248 $rm{\mu}$s. The camera-to-center distance is 926 mm, and the background is 1746 mm away.

For the diffusion flame case, due to its strong axisymmetry, only a single view is captured, and multi-view projections are synthesized assuming rotational symmetry. The same camera and lens are used. A Poisson-disk background is illuminated by a 100 W green light source (532 nm), with an exposure time of 1/3000 s. A 532 nm band-pass filter is added to suppress flame self-emission. The camera-to-center and background distances are 685 mm and 1315 mm.

\subsection{Implementation}
\label{sec:Implementation}
To facilitate reproducibility, we release the source code of this work, built upon the CT discrete reconstruction implementation of Biguri et al. (TIGRE) \cite{biguri2016} and the neural graphics primitives implementation of Tang et al. (iNGP) \cite{tang2022}, available at: https://github.com/Weihu22/Neural-Implicit-Reconstruction-for-BOS.

$\textit{Hardware.}$ The implementation of our model is based on the PyTorch framework on a workstation equipped with an Intel 13th Gen Core i9-13950HX CPU (2.20 GHz) and an NVIDIA GeForce RTX 4090 Laptop GPU.

$\textit{Encoding.}$ The Hash encoder contains 16 levels whose grid resolutions grow exponentially from 16 to 512. Each level is implemented as a trainable hash table with $2^19$ entries, where every entry stores a 2-dimensional feature vector. For a given input coordinate, its position is mapped into each level’s grid, the eight surrounding grid vertices are hashed into the table, and their features are retrieved and nonlinearly interpolated. The interpolated features from all 16 levels are finally concatenated into a 32-dimensional encoding, which serves as the input to the downstream MLP. All hash-table entries are implemented using PyTorch nn.Moduletlist function. Embedding modules and initialized with a uniform distribution in $[-1\times 10^{-4},\, 1\times 10^{-4}]$ to provide a small amount of randomness while encouraging initial predictions close to zero.

$\textit{MLP.}$ We use an MLP with two hidden layers that have a width of 64 neurons. Each hidden layer is implemented as a linear layer followed by a LeakyReLU activation with a negative slope of 0.01, and initialized using Kaiming uniform initialization. The final layer outputs a single scalar representing the refractive index, which is mapped through a $k\tan \left( x \right) + c$ activation to enforce a prescribed range [a, b]. Here, the parameters k and c are determined analytically from the interval bounds via $k + c = b$ and $- k + c = a$, and the final layer bias is initialized as $\arctan\!\left(-\frac{c}{k}\right)$ so that the initial network output corresponds to zero perturbation relative to the ambient refractive index.

This is combined with a two-stage normalization: first, the physical refractive index n is converted into a relative perturbation $\delta = n/(n_0 - 1)$, ensuring that the ambient refractive index corresponds to $\delta  = 0$; second, $\delta$ is divided by a threshold ${\delta _{{\rm{thre}}}}$, estimated from expected temperature ranges and the Gladstone–Dale relation, to confine the normalized value within a prescribed interval [a, b]. Together, this two-stage normalization and output-layer design stabilize training, improves training stability. Notably, this step is only a numerical normalization for the reconstruction algorithm. It does not change the underlying physical relationship between refractive index and temperature, nor does it overcome the weak BOS sensitivity at elevated temperatures.

$\textit{Training.}$ Our method is unsupervised and does not require ground-truth flow fields. The training data consists only of observed background displacements. For synthetic cases, all viewpoints are used for training, and the ground-truth flow field is used for validation. For real cases, all views except one are used for training, with the remaining view reserved for validation. At each iteration, we randomly sample 256 rays per image, and along each ray, up to 256 points are sampled to evaluate the network output for rendering deflection angles. Both forward and backward rendering are implemented with CUDA acceleration. The MLP weights and the hash table entries are jointly optimized using the Adam optimizer with an initial learning rate of $2 \times 10^{-2}$, $\beta = [0.9,,0.99]$, and $\epsilon = 1 \times 10^{-8}$. A cosine annealing learning rate scheduler is used to gradually reduce the learning rate from the initial value to a minimum of $1 \times 10^{-6}$ over a total of 10{,}000 iterations.

$\textit{Occupancy grids.}$ The occupancy grid is implemented as a cascaded 3D refractive-index grid to accelerate ray sampling, with the scene bounded by an axis-aligned bounding box defined as $[-2, -2, -2,, 2, 2, 2]$. The number of cascade levels is determined as $1 + \lceil \log_2(\mathrm{bound}) \rceil$, resulting in 2 cascades for our configuration. Each cascade has a fixed resolution of $128 \times 128 \times 128$. Before each epoch, the grid is updated either fully during the first 16 iterations or partially in later iterations to reduce computation. The temporary refractive-index grid is merged with the existing grid using an exponential moving average (decay = 0.95). An absolute refractive-index threshold of $1 \times 10^{-5}$ is applied to convert the grid into a bitfield representation, enabling efficient ray-marching queries.

\section{Results and discussion}
\subsection{Phantom test}
\begin{figure}[H]
    \centering
    \includegraphics[width=1\linewidth]{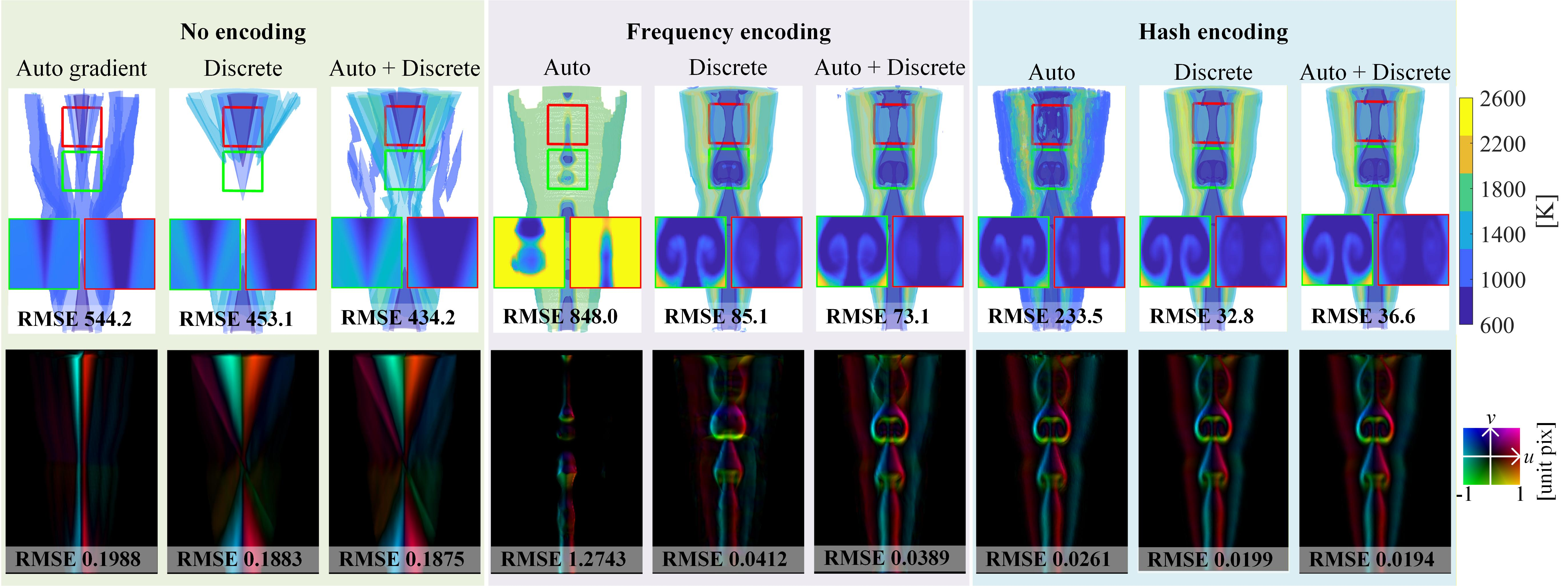}
    \caption{Comparison of reconstruction quality obtained with different encoding and gradient strategies. All configurations were trained for 10,000 iterations, varying only the input encoding and the gradient operator used for rendering. The first row presents the predicted temperature isosurfaces at 600, 1000, 1400, 1800, 2200, and 2600 K. The second row shows the predicted image-displacement colormap ($u$,$v$,1), normalized to the range [-1, 1], generated using the auto-gradient operator. RMSE represents the root-mean-square error between the predicted temperature (or displacement) and the corresponding ground truth.}
    \label{fig:autodisc}
\end{figure}

First, we compare reconstruction quality for Phantom 1 under different encodings and refractive-index differentiation strategies at the same iteration number, as shown in Fig.\ref{fig:autodisc}. Accuracy is evaluated using RMSE between predicted and ground-truth temperature (and displacement) fields. Without encoding, results show severe over-smoothing regardless of gradient strategy, with poor recovery of structural details and the worst RMSE for both temperature and displacement. Discrete and hybrid gradients provide only marginal improvement, indicating that plain MLPs are insufficient for representing highly nonlinear, multi-scale flow fields.

The Fourier frequency encoding uses 39 features with six frequency bands, improving the representation of high-frequency structures such as vortices, pulsations, and fine-scale turbulence. However, its high-frequency basis amplifies gradient noise under automatic differentiation, leading to oscillations and artifacts in temperature and displacement fields. Discrete gradients effectively suppress this noise and improve stability, while the hybrid strategy provides a balance between smooth convergence and preservation of high-frequency details, achieving the lowest RMSE and best overall reconstruction quality.

In contrast, multi-resolution hash encoding shows more balanced and robust performance across gradient strategies. It captures both large- and small-scale structures via multi-scale sparse embeddings, while avoiding amplification of gradient noise, leading to cleaner and more consistent reconstructions under Auto, Discrete, and Hybrid settings.
\end{multicols}
\vspace{-2em}

\begin{figure}[H]
    \centering
    \includegraphics[width=1\linewidth]{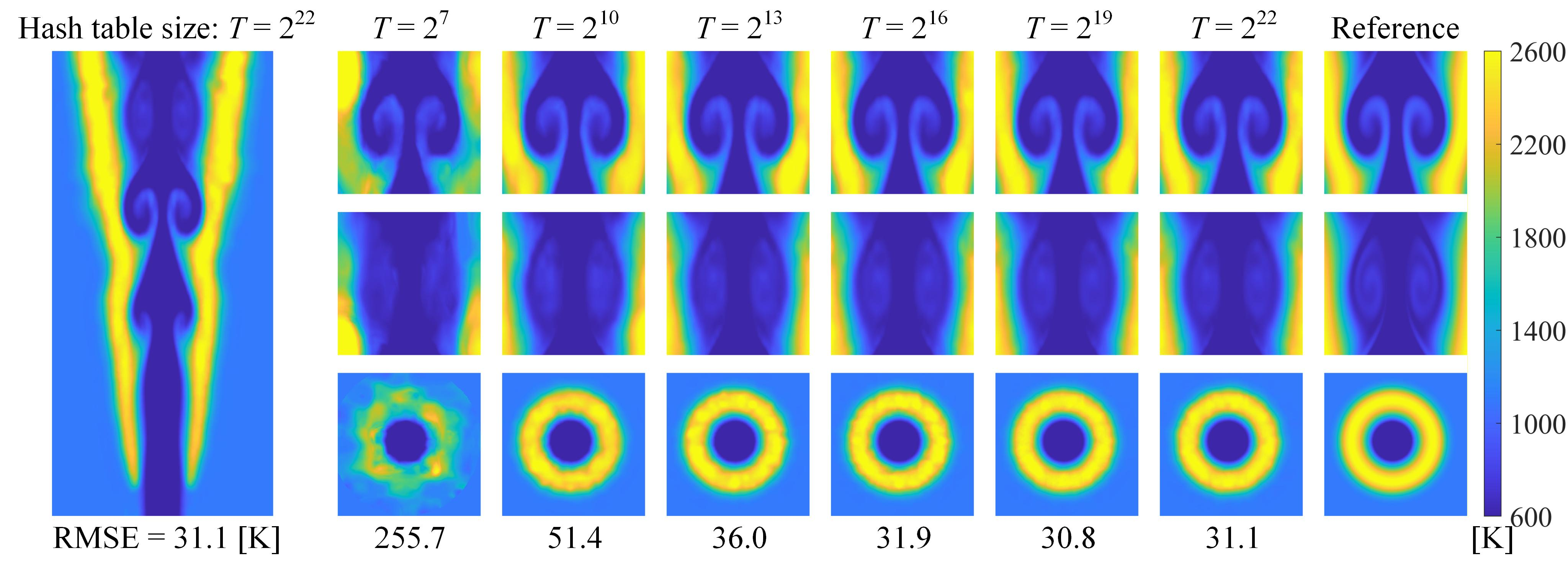}
    \caption{Approximating a flame temperature of resolution 280×588×280 with different hash table size.}
    \label{fig:hashsize}
\end{figure}

\begin{multicols}{2}

Notably, combining discrete gradients with hash encoding achieves the best balance between noise suppression and multi-scale representation, resulting in the lowest temperature RMSE and the most accurate recovery of geometry and intensity in high-gradient regions (e.g., plume boundaries with strong refractive-index variations). The hybrid gradient strategy remains stable and reasonably accurate, but does not show a clear advantage over the discrete-only strategy under hash encoding.

However, as emphasized in Section \ref{sec:Loss}, accurate automatic differentiation is essential for future physics-informed extensions, where higher-order derivatives may be required. In the present study, although the hybrid strategy slightly improves the consistency of automatically differentiated gradients, this improvement is marginal compared to the discrete-only approach, especially under hash encoding. Considering that automatic gradients are highly sensitive to noise and may destabilize training in real measurements, we adopt discrete gradients as the primary driver for network convergence, while using automatically differentiated gradients only as a monitoring signal to ensure their physical consistency. 

Fig.\ref{fig:hashsize} illustrates the reconstruction details obtained with different hash table sizes $T$. The results show that hash table size critically affects 3D flame temperature reconstruction quality. When the table is too small (e.g., $T$ = $2^7$), insufficient independent embedding slots cause severe hash collisions, leading to blurred or misidentified large-scale structures, loss of small-scale features, and pronounced artifacts and numerical diffusion. Consequently, the reconstructed temperature field deviates significantly from the ground-truth flame morphology.

As $T$ increases (e.g., $2^{10}$ to $2^{16}$), collisions decrease and spatial gradients are learned more reliably. Plume shape, shear-layer entrainment, and fine-scale turbulence become better resolved with smoother temperature isosurfaces. Increasing further (e.g., $2^{19}$ to $2^{22}$) may introduce slight degradation, indicating that hash table size effectively controls encoding resolution. Small tables suffer from embedding collisions, while sufficiently large tables enable accurate high-frequency representation. A saturation effect appears beyond $T$ = $2^{19}$, where gains diminish as most structural information is already captured, while further increasing capacity raises computational cost and requires stronger projection constraints to maintain well-posed reconstruction.
\begin{figure}[H]
    \centering
    \includegraphics[width=1\linewidth]{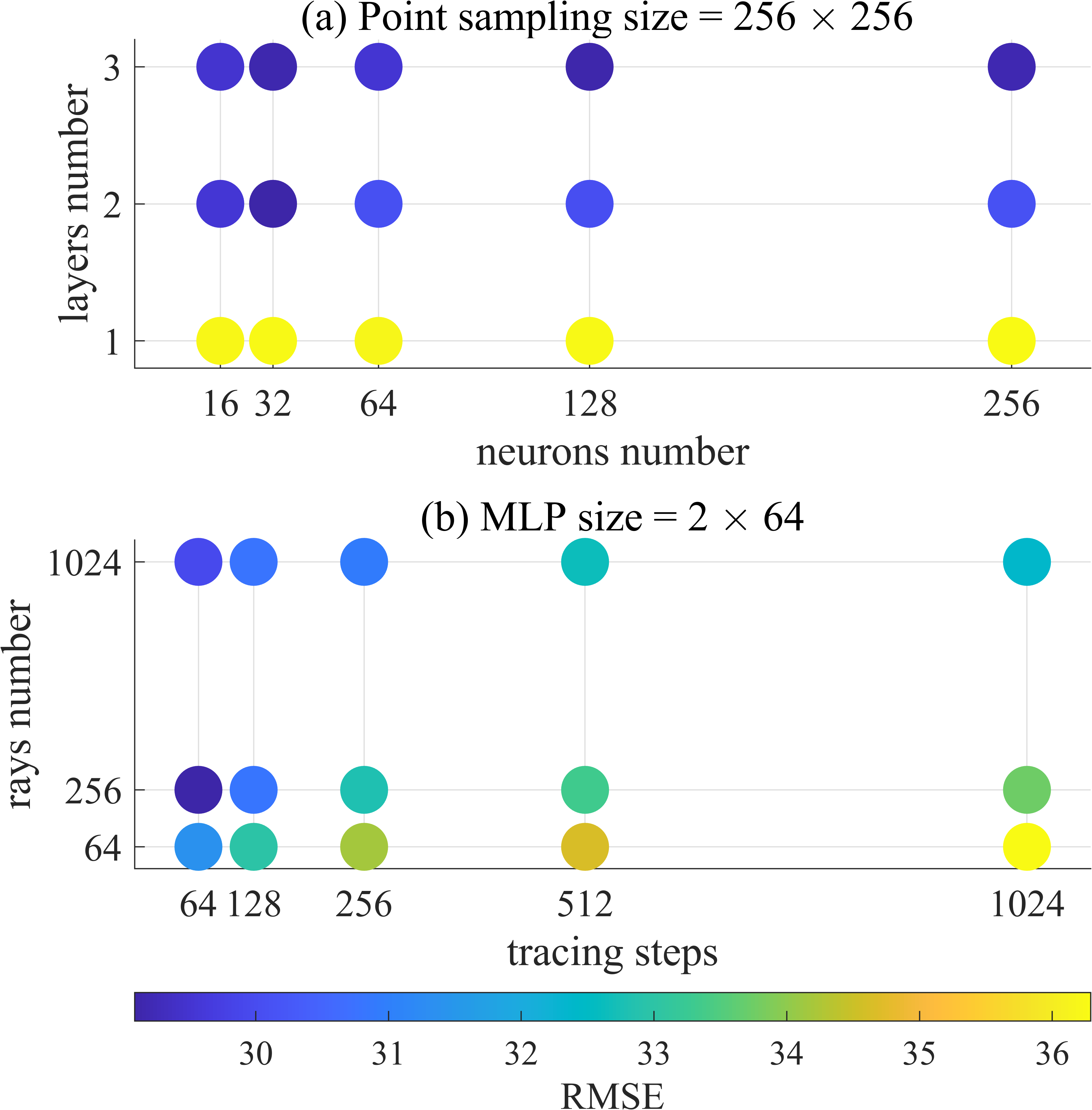}
    \caption{The effect of the MLP size (a) and the point sampling size (b) on the reconstruction quality.}
    \label{fig:paraset}
\end{figure}

To evaluate the influence of network architecture and sampling on reconstruction fidelity, we performed a parameter sensitivity study using RMSE of the reconstructed temperature field. Fig. \ref{fig:paraset}(a) shows the effect of different MLP configurations at a fixed sampling size. Network depth is the primary factor: a single-layer network underfits due to limited capacity, while deeper networks provide increased representational power. A two-layer network consistently achieves strong generalization, maintaining low RMSE even with a modest number of neurons (e.g., 16 or 32). Although a three-layer network offers comparable reconstruction accuracy, it does not yield a noticeable RMSE reduction, while incurring higher training cost. This indicates that a compact two-layer nonlinear mapping effectively captures the task’s geometric characteristics, and further increasing network capacity brings limited accuracy gains but adds computational overhead.

Fig. \ref{fig:paraset}(b) evaluates the effect of sampling size under a fixed MLP. The number of tracing steps is positively correlated with reconstruction error: the best performance occurs at 64–256 steps, while 512 or more significantly degrades accuracy, likely due to redundant, highly correlated samples that reduce optimization stability and promote overfitting to local ray-wise variations. The number of rays shows a non-monotonic trend, with the lowest RMSE at 256 rays; both under-sampling (64) and over-sampling (1024) increase errors, indicating a trade-off between information sufficiency and stability. Based on these results, we use a two-layer MLP with 64 neurons per layer and a sampling strategy of 256 rays and 256 tracing steps for final reconstruction.

\begin{figure}[H]
    \centering
    \includegraphics[width=1\linewidth]{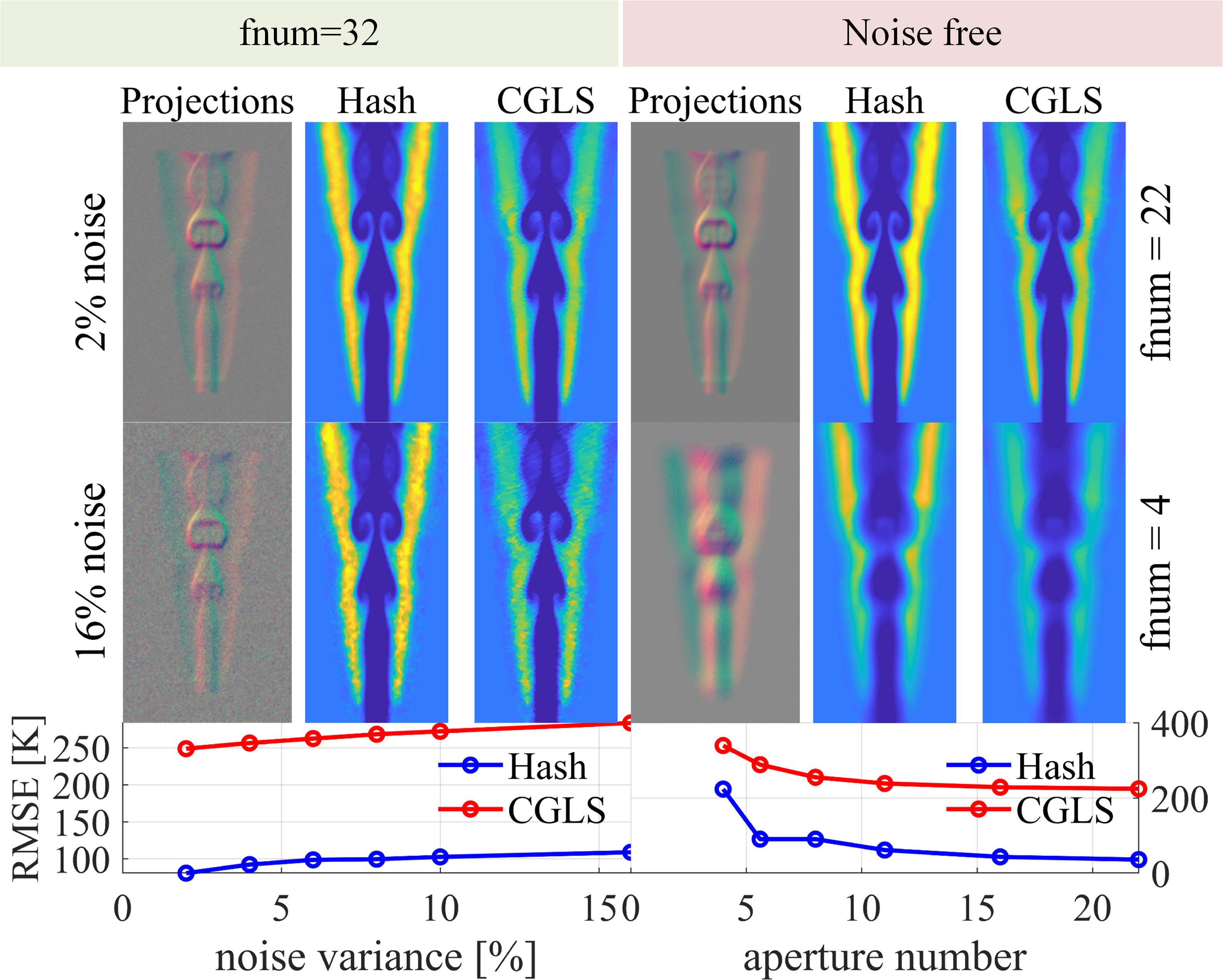}
    \caption{Error scaling with white noise variance (left three columns) and aperture number (right three columns).}
    \label{fig:noise}
\end{figure}

\begin{figure}[H]
    \centering
    \includegraphics[width=1\linewidth]{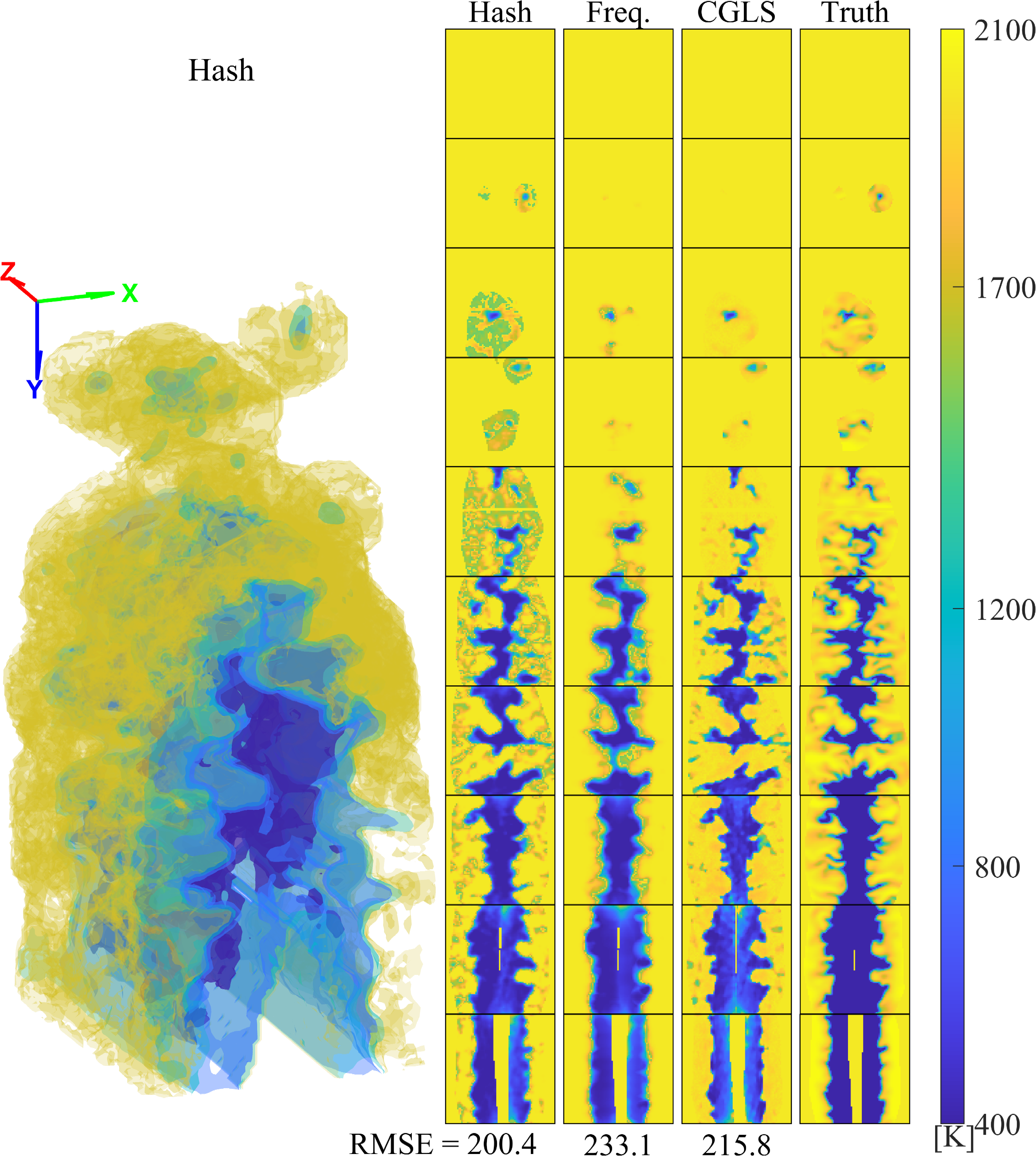}
    \caption{Temperature reconstruction results for Phantom 2 using three methods. Left: isosurfaces at values indicated by the colorbar. Right: temperature slices at different y-locations for hash encoding, frequency encoding, voxel-based CGLS, and the ground truth.}
    \label{fig:10}
\end{figure}

\begin{figure}[H]
    \centering
    \includegraphics[width=1\linewidth]{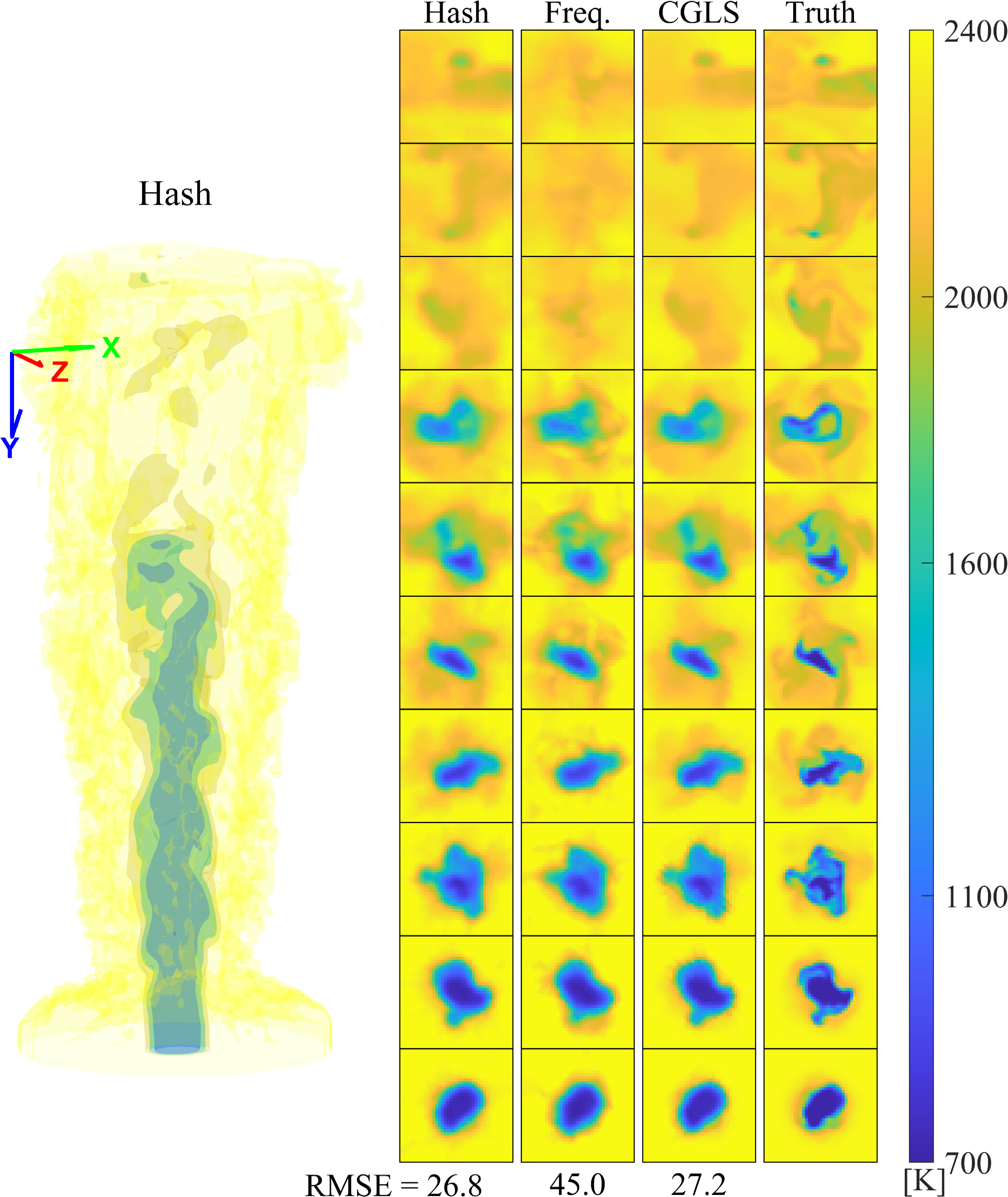}
    \caption{Temperature results for phantom 3.}
    \label{fig:11}
\end{figure}

\begin{figure}[H]
    \centering
    \includegraphics[width=1\linewidth]{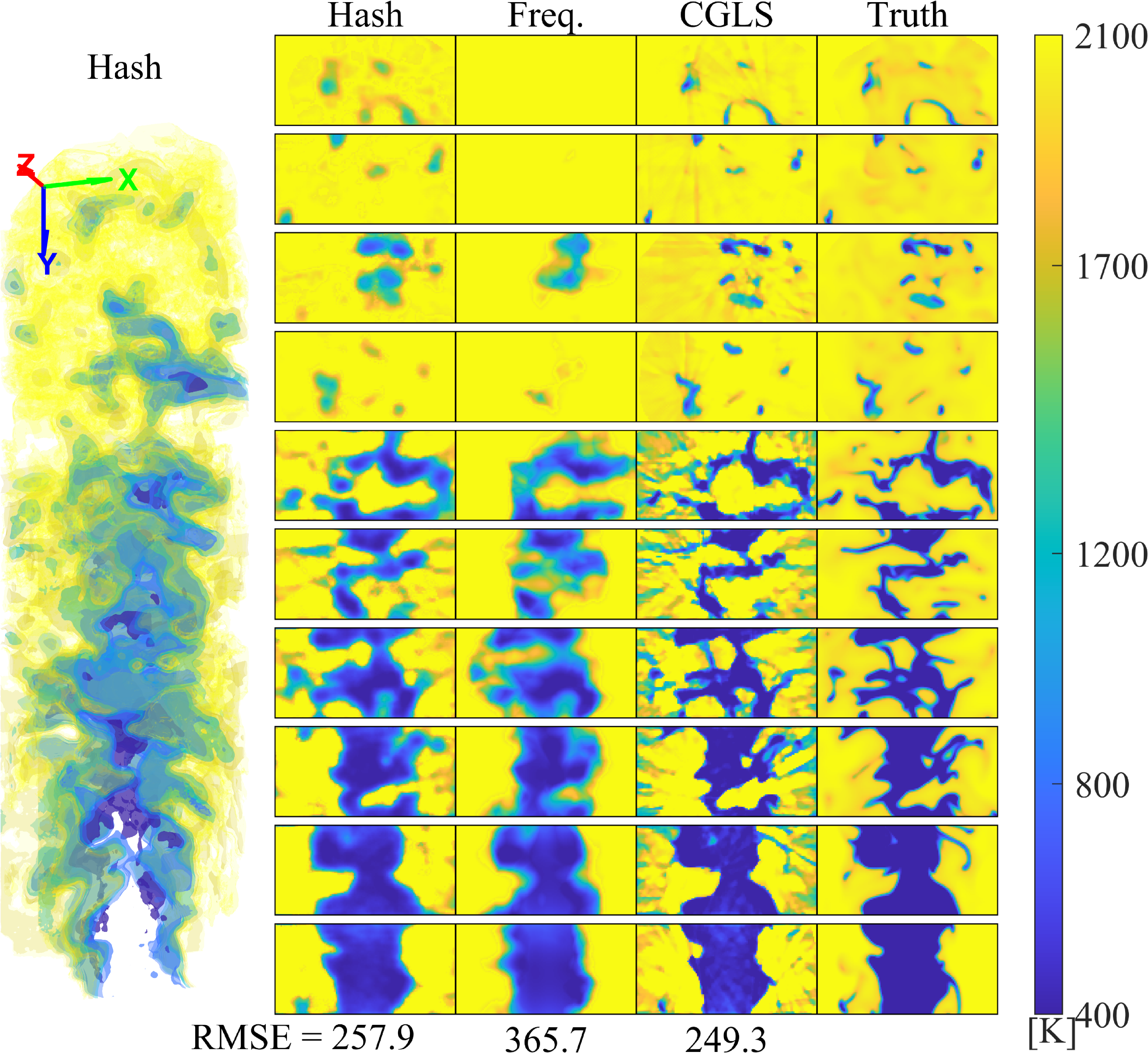}
    \caption{Temperature results for phantom 4.}
    \label{fig:12}
\end{figure}

\begin{figure}[H]
    \centering
    \includegraphics[width=1\linewidth]{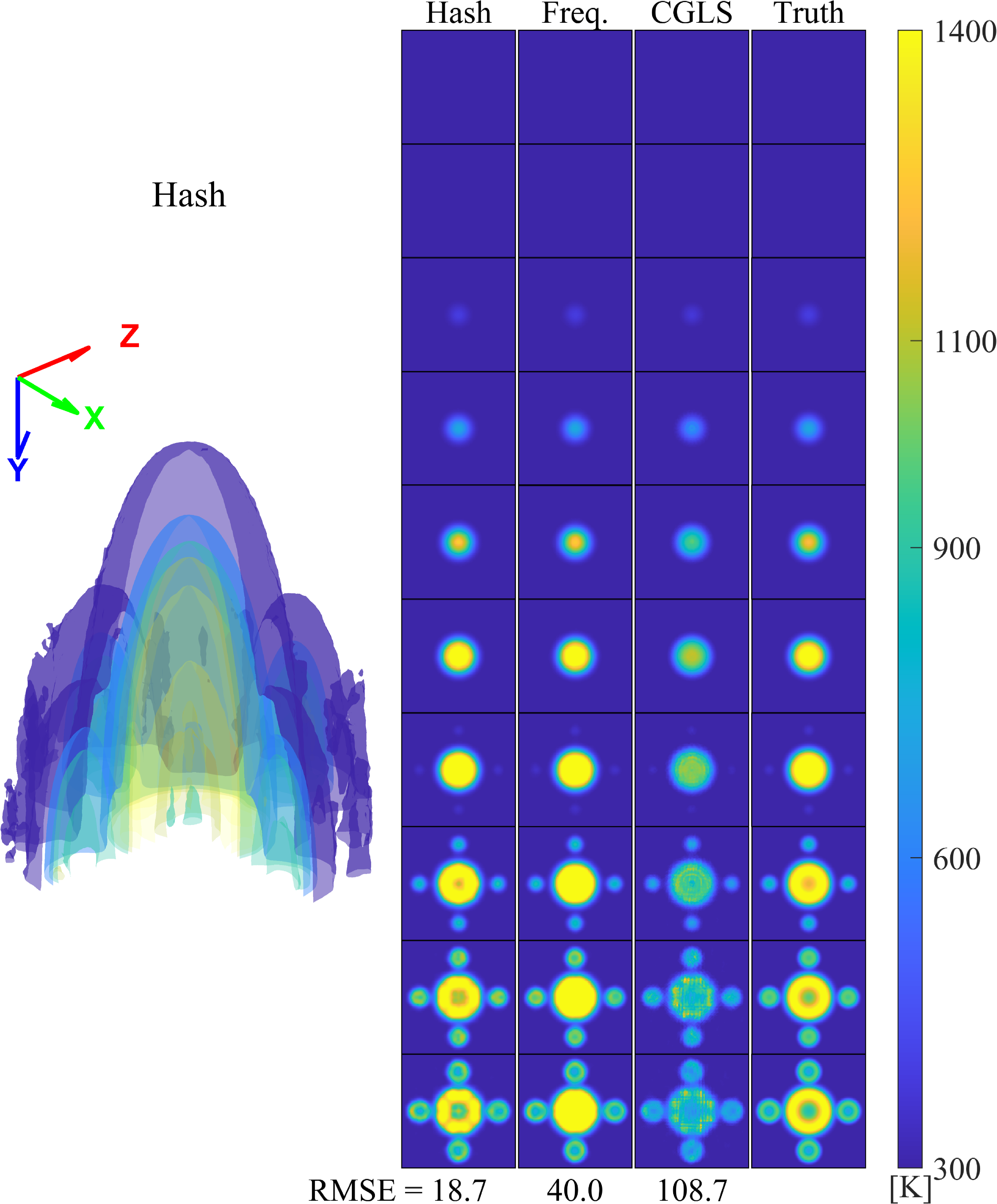}
    \caption{Temperature results for phantom 5.}
    \label{fig:13}
\end{figure}

\begin{figure}[H]
    \centering
    \includegraphics[width=1\linewidth]{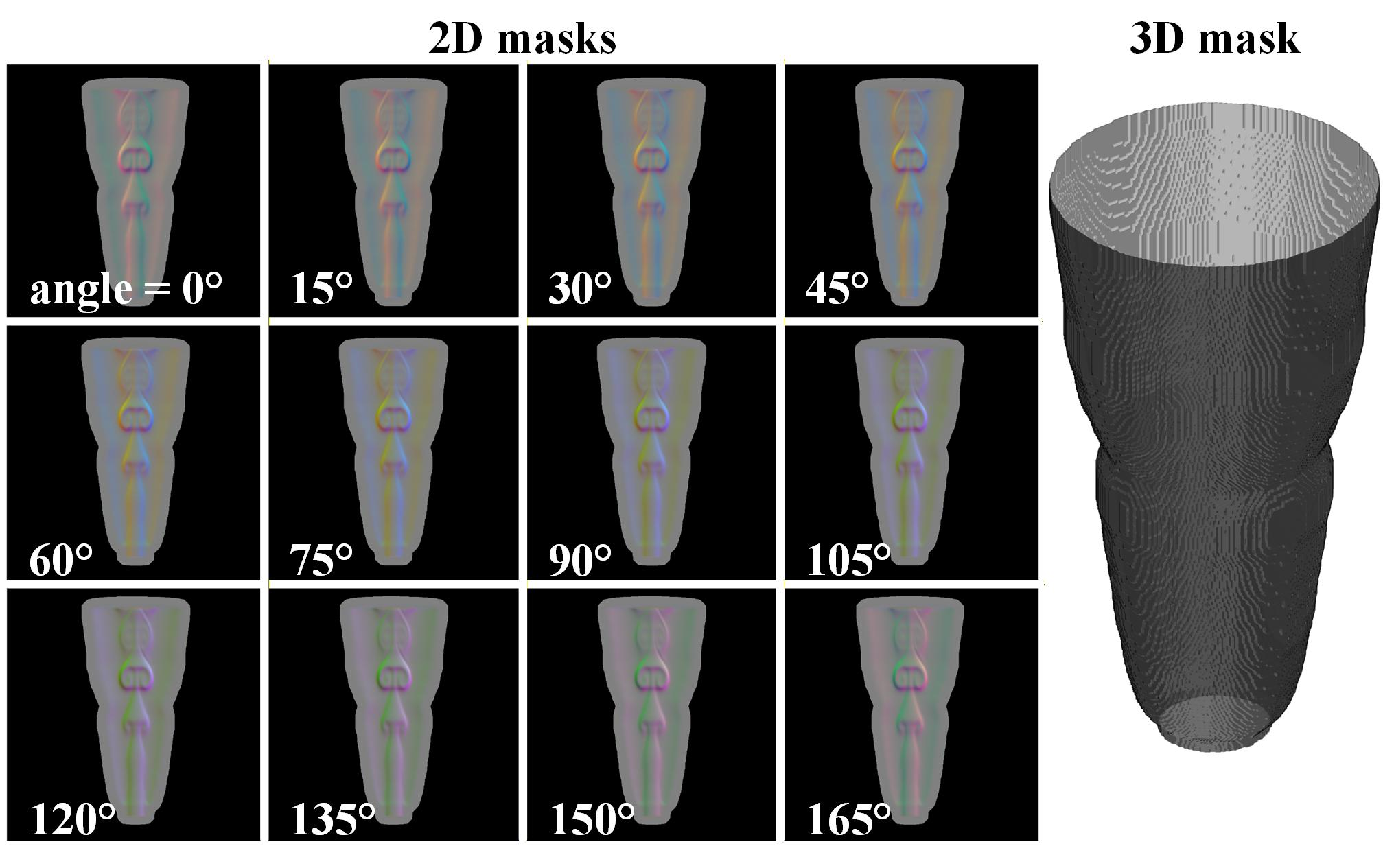}
    \caption{2D (left) and 3D (right) masks of for phantom 1.}
    \label{fig:14}
\end{figure}

Fig. \ref{fig:noise} evaluates robustness under noise (left) and defocus blur (right), comparing the proposed hash-encoding reconstruction (Hash) with voxel-based CGLS (Conjugate Gradient Least Squares). The left panel varies noise variance at a fixed aperture (32), while the right varies aperture under noise-free conditions. As noise increases, both methods show higher RMSE, but Hash consistently remains lower (below 150), indicating stronger noise resistance. With increasing aperture, both methods improve due to reduced blur, while Hash maintains lower RMSE across all cases. Even at noise levels above 10\%, its RMSE stays below 100, showing stable performance under degraded conditions. Visual comparisons further confirm that Hash preserves clearer structures and finer details, whereas CGLS exhibits more noticeable degradation. Overall, the proposed method demonstrates superior robustness to both noise and defocus blur.
\vspace{-2em}
Fig. \ref{fig:10}-Fig. \ref{fig:13} further evaluate our model by applying it to flame phantoms exhibiting different turbulence characteristics, and compares the reconstruction quality against the existing frequency-encoding neural implicit reconstruction method as well as the voxel-based CGLS algorithm. 
The frequency-encoding network alleviates these discretization artifacts to some extent but exhibits a distinct low-pass filtering effect. This property enables it to achieve relatively lower RMSE on Phantom 5, whose temperature field varies smoothly and resembles a laminar flame. However, for phantoms dominated by sharp gradients and turbulent structures (e.g., Phantom 2, Phantom 3 and Phantom 4), the frequency-encoding approach tends to over-smooth fine-scale features, leading to the suppression of small vortex structures and attenuation of peak temperature regions.
The proposed hash-encoding method shows consistently strong performance across Phantoms 2, 3, and 5, achieving the best overall reconstruction quality. It suppresses noise while preserving key flow structures, enabling clearer recovery of entrainment and asymmetric temperature features. This advantage is most evident in Phantom 5, where it cleanly separates multiple flame sources with minimal cross-talk, outperforming frequency encoding and CGLS. In Phantom 4, although CGLS achieves slightly lower RMSE, it introduces noticeable peripheral artifacts, while the hash result remains cleaner and more spatially consistent, indicating better robustness.
Finally, we compare reconstruction quality with and without the mask. The mask is shown in Fig. \ref{fig:14} and errors are summarized in Table \ref{tab:mask}. Results show that masking constrains the solution space, focusing the network on effective flow regions and reducing background interference. This improves convergence stability and reconstruction accuracy, indicating the benefit of the masking strategy.

\end{multicols}
\begin{table}[H]
\centering
\caption{Reconstruction temperature error (unit K) using hash encoding with a 3D mask versus without a 3D mask, based on 2D mask ray sampling. One epoch corresponds to a complete pass over the training projection data and represents 12 iterations in our case.}
\label{tab:mask}

{\fontsize{10pt}{11pt}\selectfont
\renewcommand{\arraystretch}{1.25}
\setlength{\tabcolsep}{4pt}

\resizebox{0.95\textwidth}{!}{
\begin{tabular}{c|cc|cc|cc|cc|cc}
\hline
\multirow{2}{*}{Epoch} 
& \multicolumn{2}{c|}{Phantom 1}
& \multicolumn{2}{c|}{Phantom 2}
& \multicolumn{2}{c|}{Phantom 3}
& \multicolumn{2}{c|}{Phantom 4}
& \multicolumn{2}{c}{Phantom 5} \\
\cline{2-11}

& With & Without
& With & Without
& With & Without
& With & Without
& With & Without \\
\hline

50  & 47.3 & 54.5 & 189.0 & 168.0 & 28.1 & 164.5 & 277.7 & 305.1 & 26.6 & 52.1 \\
150 & 34.7 & 34.3 & 191.0 & 185.9 & 26.9 & 161.1 & 265.3 & 338.5 & 21.6 & 48.1 \\
250 & 33.6 & 31.6 & 200.0 & 207.2 & 26.6 & 157.7 & 261.8 & 373.1 & 19.6 & 52.2 \\
350 & 34.0 & 31.4 & 201.4 & 219.3 & 26.8 & 156.8 & 259.3 & 395.4 & 19.1 & 50.1 \\
450 & 33.2 & 31.0 & 198.3 & 228.7 & 26.7 & 154.2 & 257.9 & 417.0 & 18.8 & 47.3 \\
550 & 33.4 & 31.7 & 199.7 & 237.7 & 26.9 & 154.3 & 259.9 & 430.7 & 18.7 & 47.4 \\
650 & 32.5 & 31.7 & 200.6 & 246.0 & 26.8 & 153.6 & 257.4 & 437.3 & 18.8 & 47.3 \\
750 & 33.0 & 31.7 & 200.6 & 247.1 & 26.8 & 153.0 & 258.4 & 440.0 & 18.7 & 46.9 \\
834 & 32.8 & 31.6 & 200.4 & 249.2 & 26.8 & 153.1 & 257.9 & 442.8 & 18.7 & 47.0 \\
\hline

\end{tabular}
}
}
\end{table}

\begin{multicols}{2}

\subsection{Experiment test}
\begin{figure}[H]
    \centering
    \includegraphics[width=1\linewidth]{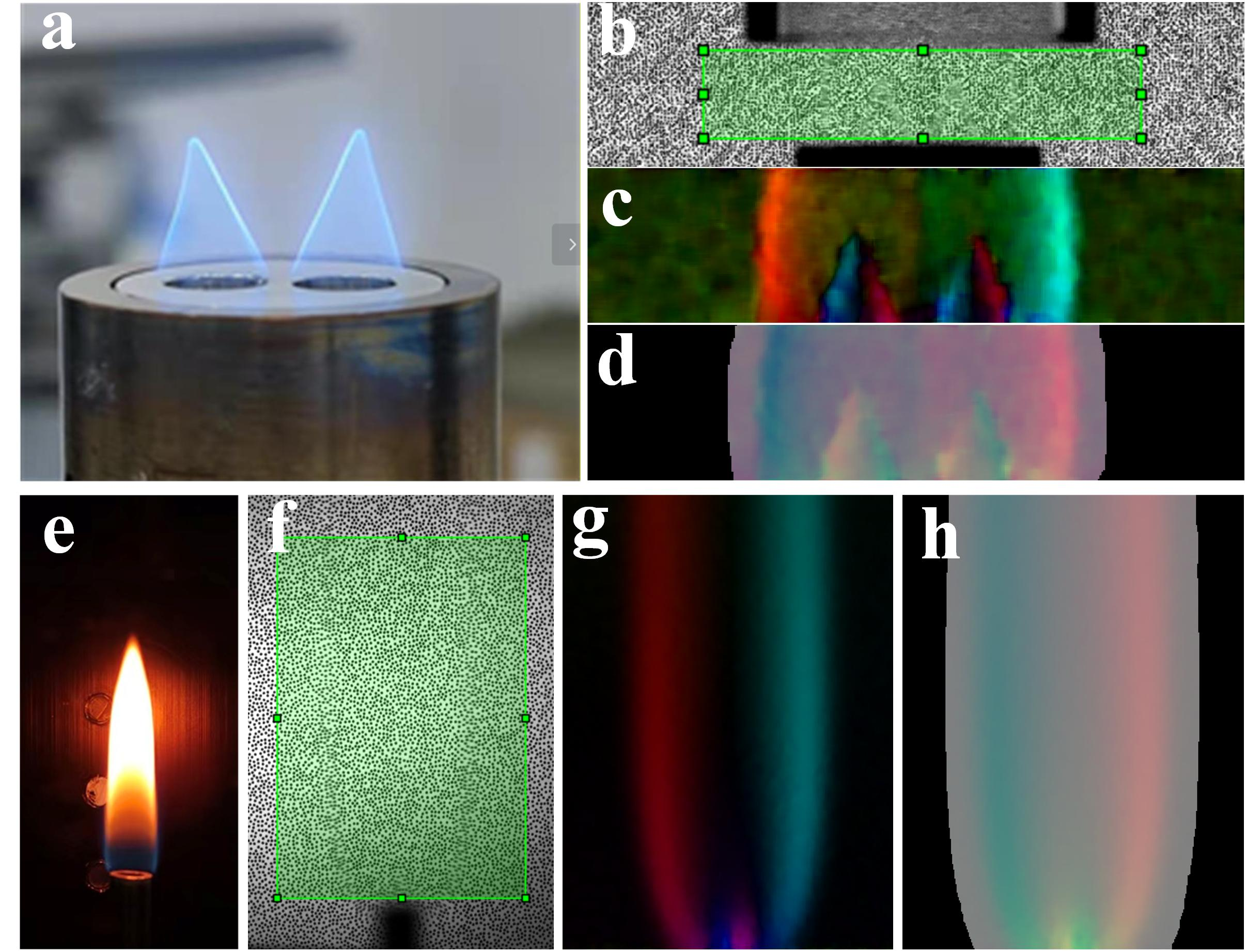}
    \caption{Measurement results of background schlieren of double flame (a)-(d) and diffusion flame (e)-(h): (a)\&(e) Image of the flames; (b)\&(f) Background images post-processed via contrast enhancement (the green box denotes the region of interest); (c)\&(g) Image displacement maps ($u$,$v$,1) in the region of interest; (d)\&(h) Deflection-angle maps $({\varepsilon _x},{\varepsilon _y},{\varepsilon _z})$ incorporating a 2D mask.}
    \label{fig:15-flames}
\end{figure}

Fig. \ref{fig:15-flames} presents the measurement results for two real flame cases. The double flame (a)–(d) is a premixed flame characterized by a relatively high temperature and a strongly asymmetric spatial distribution. Due to its small physical size in the experiment, measurement noise becomes more pronounced, which is reflected in the displacement and deflection results shown in (c) and (d). In contrast, the diffusion flame (e)–(h), associated with less complete combustion and a comparatively lower temperature, exhibits a larger spatial extent. As a result, the influence of measurement noise is less significant, leading to more stable observations.

For both cases, the first nine views (0–144°) are used for training, while the tenth view (162°) is reserved for evaluation. The reported results correspond to the comparison between the observed and predicted image displacement at this held-out viewing angle, using RMSE and peak signal-to-noise ratio (PSNR) as quantitative metrics for assessing reconstruction performance. The double flame is reconstructed at a resolution of 176 × 56 × 176, while the diffusion flame is reconstructed at 124 × 188 × 124, both with a voxel size of 0.25 × 0.25 × 0.25 $mm^3$. 

\begin{figure}[H]
    \centering
    \includegraphics[width=0.75\linewidth]{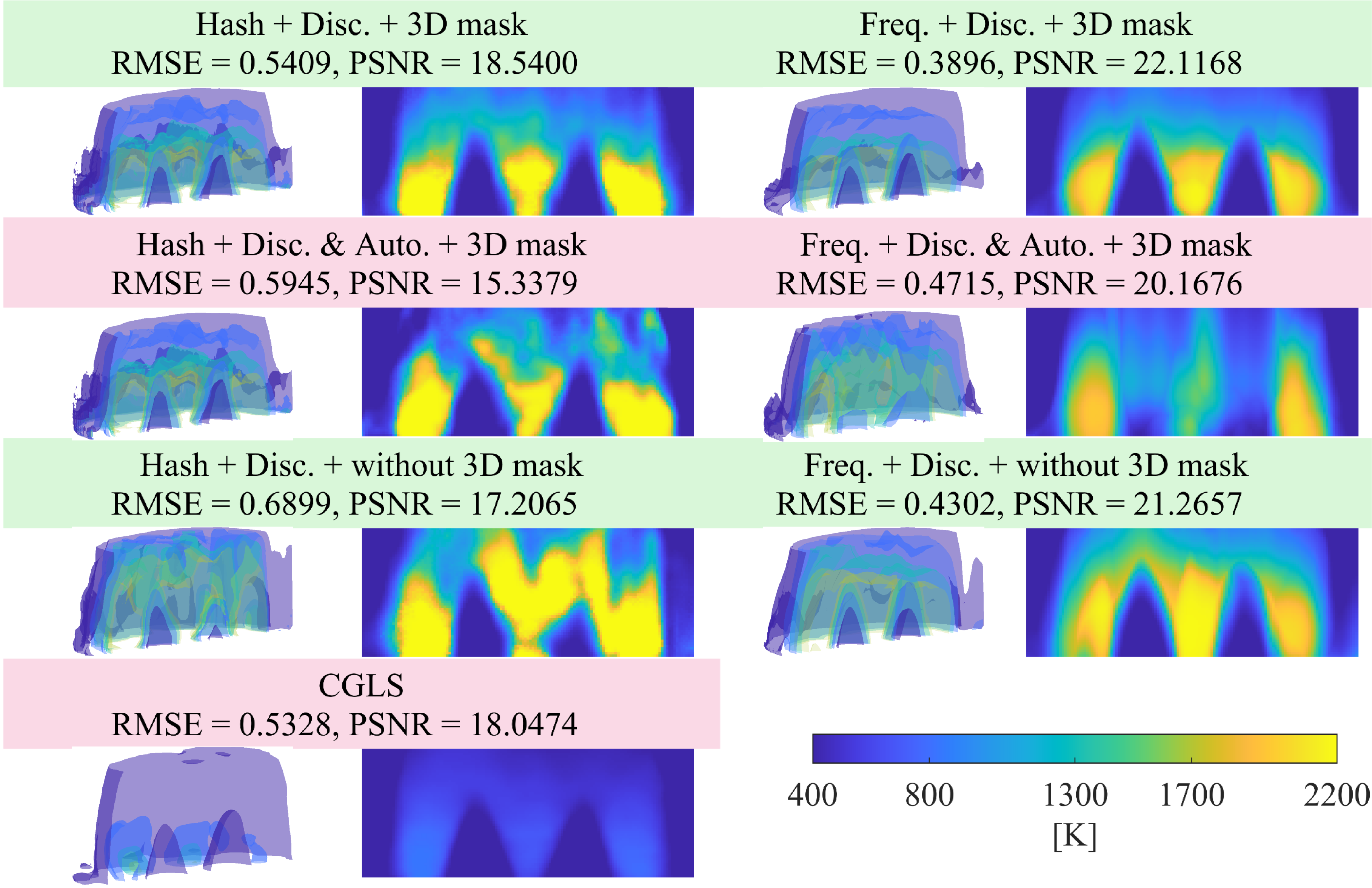}
    \caption{3D and 2D slices of the reconstructed double flame temperature field using different reconstruction configuration.}
    \label{fig:16}
\end{figure}

\begin{figure}[H]
    \centering
    \includegraphics[width=1\linewidth]{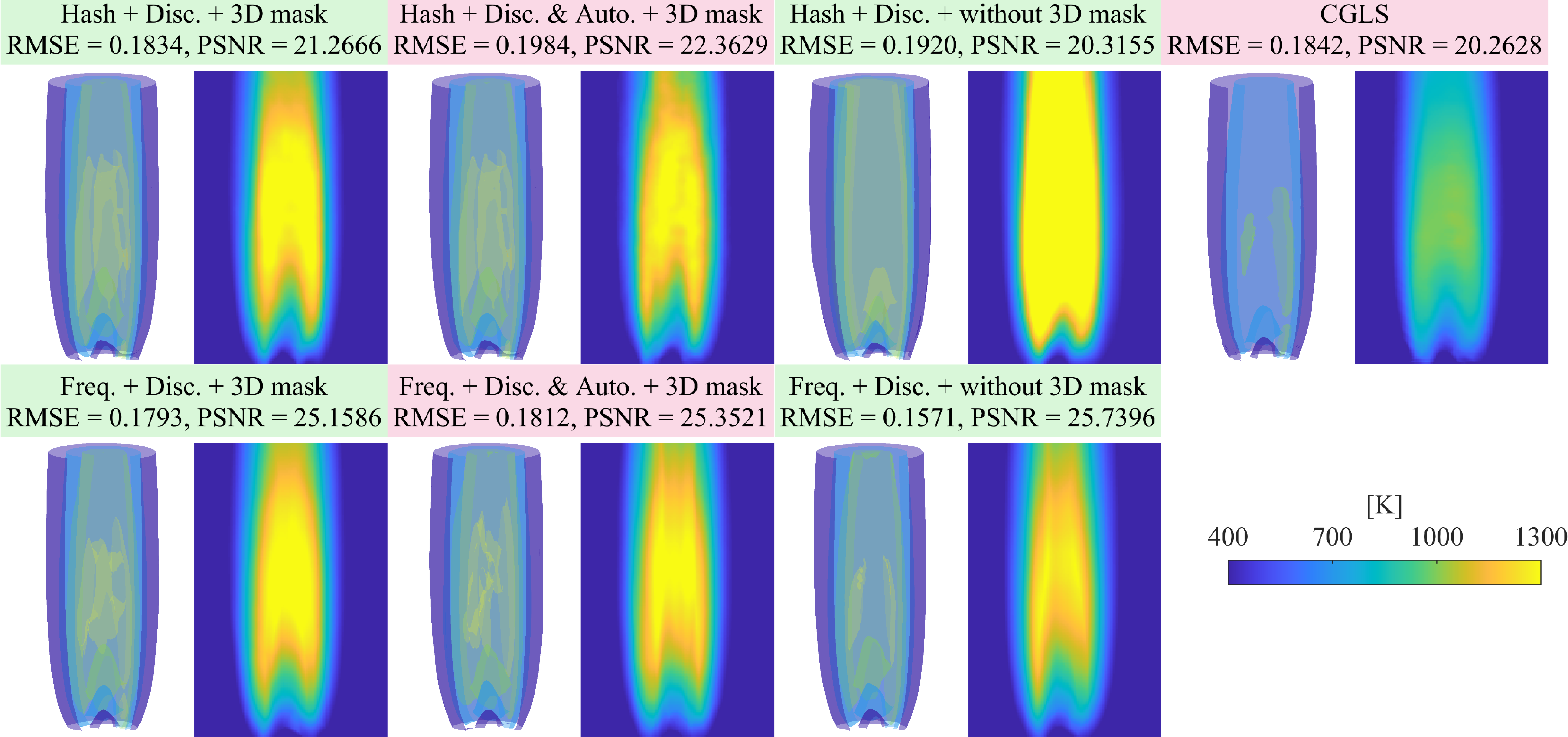}
    \caption{3D and 2D slices of the reconstructed diffusion flame temperature field using different reconstruction configuration.}
    \label{fig:17}
\end{figure}

The reconstruction results for the dual flame and diffusion flame are presented in Fig. \ref{fig:16} and Fig. \ref{fig:17}. The prior temperature is constrained to 300–2200 K during training. In both cases, it is observed that the voxel-based CGLS method struggles to obtain a valid solution at high resolution, as adjusting global smoothness alone is insufficient to regularize the inverse problem effectively.

In the double-flame case, the frequency encoding with discrete gradient loss and a 3D mask achieves the best RMSE and PSNR, producing more physically consistent temperature fields with clear spatial gradients. In contrast, hash-based methods exhibit stronger space fitting ability and yield flame front structures that are more closely aligned with the observed BOS deflection patterns. However, they are more prone to overfitting noisy measurements, leading to non-physical saturation near the flame root, where temperatures collapse toward 2200 K and gradients nearly vanish. This indicates optimization saturation, likely caused by reduced refractive-index sensitivity at high temperatures and accumulated reconstruction errors. Frequency encoding mitigates this issue by providing a smoother, globally continuous representation with implicit regularization, which helps suppress overfitting to high-temperature plateaus. Meanwhile, both removing the 3D mask and using auto-differentiated gradients destabilize convergence under noisy observations.

For the diffusion flame, which has lower temperature and lower measurement noise, both hash encoding and frequency encoding with discrete gradient loss and a 3D mask perform similarly, with temperatures converging to around 1300 K. Frequency encoding achieves slightly better RMSE and PSNR. Without the 3D mask, hash encoding again exhibits large saturated high-temperature regions, while frequency encoding yields better metrics but introduces unnaturally sharp structures, indicating reduced physical consistency; thus, the 3D mask remains preferable. In addition, the auto-differentiated gradient loss does not improve agreement with observed image shifts.
\end{multicols}

\clearpage
\begin{multicols}{2}

\section{Conclusion}
This paper presents a compact neural refractive-index-primitive framework for BOS tomography, where the flow field is modeled as a continuous 3D refractive-index distribution parameterized by an MLP. The method integrates different encoding strategy (hash and frequency), various gradient computation (automatic differentiation and central differences), a 3D flow mask, and a ray-sampling scheme to improve convergence stability, reduce artifacts, and enhance reconstruction accuracy.

Extensive phantom and real flame experiments demonstrate consistent performance across different conditions. Hash encoding provides strong multi-scale representation and accurate reconstruction in relatively clean cases, while frequency encoding offers greater stability under noisy measurements. Discrete gradient supervision and the 3D mask further improve physical consistency and reduce artifacts. 

Overall, the work focuses on improving refractive-index reconstruction accuracy, with temperature recovery being an indirect consequence under simplified assumptions. Stronger combustion modeling would require additional physics constraints such as data assimilation or coupled thermo-chemical modeling, which are left for future work.\\

\noindent\textbf{CrediT authorship contribution statement}
\textbf{Xinyi Lu:} Writing – original draft, Validation, Investigation, Formal analysis. \textbf{Wei Hu:} Writing –review \& editing, Methodology, Conceptualization, Supervision. \textbf{Zizhou Liao:} Data curation. \textbf{Zheng Wang:} Methodology. \textbf{Yue Zhang:} Visualization. \textbf{Jingxuan Li:} Writing –review \& editing, Supervision.\\

\noindent\textbf{Declaration of competing interest}
The authors declare that they have no known competing financial interests or personal relationships that could have appeared to influence the work reported in this paper.\\

\noindent\textbf{Acknowledgments}
This work was supported by Zhejiang Provincial Natural Science Foundation of China [grant number LQN25E060008], Ningbo Yongjiang Talent Programme [grant number 2024A-416-G] and Undergraduate Innovation and Entrepreneurship Training Program [grant number S202510006191].\\

\bibliographystyle{elsarticle-num}
\bibliography{refs}

\end{multicols}

\end{document}